\documentclass[preprint,12pt]{elsarticle}

 \newcommand*\diff{\mathop{}\!\mathrm{d}}

\usepackage[utf8]{inputenc} 
\usepackage[english]{babel}
\usepackage{afterpage}
\usepackage[perpage]{footmisc}

\usepackage{quoting}
\quotingsetup{font=small}
\usepackage{emptypage}

\usepackage{pgfplots}
\pgfplotsset{width=7cm,compat=1.8}

\usepackage{amsfonts}
\usepackage{amsmath}
\numberwithin{equation}{section}
\usepackage{color}

\usepackage{hyperref}
\usepackage{textcomp}
\usepackage{gensymb}

\usepackage{float}
\usepackage[caption = true]{subfig}
\usepackage{booktabs}
\usepackage{siunitx}

\usepackage{amsthm}
\makeatletter
\def\th@plain{%
	\thm@notefont{}
	\itshape 
}
\def\th@definition{%
	\thm@notefont{}
	\normalfont 
}
\theoremstyle{plain}
\newtheorem{thm}{Theorem}[section]

\theoremstyle{definition}
\newtheorem{defn}{Definition}[section]

\theoremstyle{remark}

\renewcommand{\vec}{\boldsymbol}
\usepackage[nolist]{acronym}

\usepackage{caption}
\usepackage{graphicx}
\usepackage{amssymb}

\usepackage{lineno}

\journal{Biosystems}

\begin{document}

\begin{frontmatter}


\title{A study of dependency features of spike trains through copulas}


 \author[label1, label3]{Pietro Verzelli }
 \author[label2]{Laura Sacerdote }
 \address[label1]{Università della Svizzera italiana, Lugano, Switzerland.}
 \address[label2]{Università degli studi di Torino, Turin, Italy}
 \address[label3]{Referring author: pietro.verzelli@usi.ch}

\begin{abstract}
Simultaneous recordings from many neurons hide important information and the connections characterizing the network remain generally undiscovered despite the progresses of statistical and machine learning techniques. Discerning the presence of direct links between neuron from data is still a not completely solved problem. \\
To enlarge the number of tools for detecting the underlying network structure, we propose here the use of copulas, pursuing on a research direction we started in  \cite{sacerdote2012detecting}. 
 Here, we adapt their use to distinguish different types of connections on a very simple network. Our proposal consists in choosing suitable random intervals in pairs of spike trains determining the shapes of their copulas. We show that this approach allows to detect different types of dependencies. \\
We illustrate the features of the proposed method on synthetic data from suitably connected networks of two or three formal neurons directly connected or influenced by the surrounding network. We show how a smart choice of pairs of random times together with the use of empirical copulas allows to discern between direct and un-direct interactions.

\end{abstract}


\end{frontmatter}



\section{Introduction}

Improved measurement devices collect data from an increasing number of neurons and disclose new opportunities of learning from complex data sets. However, despite the high technology level of these devices, the structure of the observed network remains hidden.  The old problem of “elucidate the representation and transmission of information in the nervous system” \cite{perkel1968neural} is still open despite the important progresses determined by statistical and machine learning methods to guess the structure of the network (see \cite{kass2018computational} for a recent review). 
Different types of neuronal data request the use and the development of specific statistical tools. We focus here with those related with neuronal spiking activity. Parallel recordings from neurons hide the structure of the network generating the data. Suitable statistical methods may help to guess connections between the neurons. There exist many statistical techniques for the analysis of massively parallel spike trains (see \cite{kass2018computational}). When the focus is on the structure of small networks, some methods involve generalized linear models \cite{brillinger1988maximum},  \cite{stevenson2009bayesian} with their difficulties \cite{eldawlatly2009identifying} or variants of the Cox method \cite{cox1972multivariate}, \cite{borisyuk1985new}, \cite{masud2011statistical}. 
Existing studies often involve correlation measures. However, correlation recognizes mainly linear dependencies and part of the information contained in the data is wasted with these approaches. Furthermore, the analysis  may reveal dependencies without suggesting any structure of the underlying network. Hence, it is desirable to improve methods to discern different features corresponding to different types of dependencies between \acp{ISI}. \\

Roughly speaking, two main causes determine dependencies between \acp{ISI} of two spike trains: either  the two neurons are directly connected (direct dependence) or both the neurons receive their input by a common network (indirect dependence). A mixing of these two features is surely possible but we disregard this possibility in this paper. To guess the network structure requests methods to recognize direct neuronal links. Furthermore, when a direct connection between the neurons exists, data should allow to recognize pre- from post-synaptic neurons. To this aim, inspired by a previous work \cite{sacerdote2012detecting}, we investigate the use of copulas \cite{sacerdote2010copulas}: their use allows one to detach the information about the joint behavior from the marginal distributions. For this reason, their use in the context of the analysis of neural data is increasing in popularity in recent years \cite{berkes2009characterizing, hu2014copula, hu2015copula, hu2016joint, onken2016mixed}.
We consider two types of problems: the dependencies between  \acp{ISI} of pairs of neurons and the dependencies between spike trains.  Our goal is to recognize specific features in copulas of suitable chosen pairs of time intervals associated to directly or indirectly connected neurons. We show through examples that the copulas associated to neurons characterized by direct connections  differ from those associated to neurons receiving their input by a common network. Furthermore, we  introduce a method based on forward and backward times that allows to recognize pre- from post-synaptic neurons.

We illustrate the power of  methods based on copulas  on synthetic data. We simulate networks of two or three neurons to get parallel spike trains. We use a three dimensional \ac{LIF} model with correlated noise to simulate the effect of the surrounding network on the single neuron. To reproduce the effect of direct connection between neurons we start from independent \ac{LIF} models. Then we introduce the dependence between the spike activities by assuming that the firing of a neuron determines a jump in the membrane potential of other neurons. Positive or negative jumps are used to reproduce excitation or inhibition. We developed an open source software \ac{NERVE}%
\footnote{The complete code and a short guide on how to use it to replicate the experiments can be found at:
https://github.com/verzep/NERVE}
for such simulations and we studied the copulas of interest making use of open source packages in R.

The systematic study of different copulas associated to the network allows to recover the simulated network. Experimental data are surely more complex but the present study encourages to devote new energies to a systematic study of possible copulas shapes determined by other types of inter-neural connections.

The paper is organized as follows. In Section~\ref{sec:copulas} we recall the definition of Copula and Empirical Copula. Furthermore, we report the celebrated  Sklar's theorem \cite{sklar} expressing multivariate cumulative distributions in terms of copulas. In Section \ref{sec:dependencies} we introduce a set of pairs of random times characterizing the spike trains.  Following \cite{sacerdote2012detecting}, we introduce forward times. In \ref{sec:dependencies}, we propose the comparison of copulas based on forward and backward times to investigate the types of connections in the network.
In Section \ref{sec:neural_models} we briefly present the two neuronal models used in \cite{sacerdote2012detecting} that we apply to generate synthetic data from networks of neurons exhibiting direct or indirect connections. We also introduce a generalization of such models for the case of networks of three neurons.
Finally, in Section \ref{sec:results} we present examples of application of the proposed method to synthetic data obtained from networks of two or three neurons with prescribed connections. A final Section reports conclusions and suggestions for future studies.

\section{Copulas} \label{sec:copulas}

\subsection{Preliminary definitions}\label{subsec:preliminary_defs}
In this Section we limit ourselves to define Copulas and to report Sklar's theorem. For a more complete study on the subject we refer to the classical text \cite{nelsen2006introduction} or to the more recent book \cite{durante2015principles}.
Heuristically, a copula is a function that joins multivariate distribution functions to their one dimensional marginal distribution, i.e it allows to separate the marginal  from the joint contribution in a joint distribution. More formally, Nelsen \cite{nelsen2006introduction} gives the following definition for the $2$-dimensional case:

\begin{defn}[Copula] \label{def:2copula}
A 2-d copula is a function $C \colon [0,1]^{2} \to \mathbb[0,1]$ with the following properties:
\begin{enumerate}
		\item For each $ u,v$ in $[0,1]$:
		\begin{subequations}
			\begin{align*}
				C(u,0) =0 \quad C(0,v) = 0  \\					
				C(u,1)= u \quad C(1,v)=v 
			\end{align*}
		\end{subequations}
		\item For each $ v_1,v_2,u_1,u_2$ in $[0,1]$  such that $ v_1 < v_2, u_1<u_2$:
			\begin{equation*} \label{eqn:copdef:inc}
			C(u_2,v_2) - C(u_2, v_1) - C(u_1,v_2) +C(u_1,u_2) \ge 0
			\end{equation*}		
\end{enumerate} 
\end{defn}

The benefit of copulas in statistical analysis is related with Sklar's theorem.

\begin{thm}[Sklar's]
	\label{thm:sklar}
	Let $H$ be a joint distribution, with margins $F$ and $G$. Then there exists a copula $C$ such that for all $x,y \in \overline{\mathbb{R}}$
	\begin{equation*} \label{eqn:sklar}
	H(x,y) = C \left( F(x), G(y)\right)
	\end{equation*}
	if $F$ and $G$ are continuous, $C$ is unique. 
\end{thm}
This theorem holds even for the multivariate case.

Copulas capture all the information related to the joint behavior (i.e., dependencies) of a multivariate random variable, but without involving the marginal distributions. An important property for our aims is that they also catch non-linear information.

Closed form expressions for copulas are known in a limited number of instances. Copulas characterized by closed form expression are divided in parametric families (e.g., Gaussian, Archimedean, see \cite{nelsen2006introduction} for an extensive discussion). 
A copula that plays a special role is the \emph{independent copula} defined as:
\begin{defn}[Independent copula]\label{def:ind_cop}
Given $u,v \in [0,1]$ we define the \emph{independent copula} as
	\begin{equation}
		\Pi(u,v) := uv
	\end{equation}
\end{defn}

The following theorem describes the relation between the probabilistic concept of independence and the independence copula defined in section~\ref{def:ind_cop}.
\begin{thm}
	Let $X$ and $Y$ be \acp{RV} with joint \ac{CDF} $H$. They are independent if and only if the copula of $H$, denoted with $C_H$, is identically equal to the independent copula, i.e.,:
	\begin{equation}
		C_H(u,v) = \Pi(u,v) 
	\end{equation}
\end{thm}

\subsection{Empirical copulas}\label{subsec:empirical}

Copulas can be estimated from a dataset using \emph{pseudo-observations}, that are usually represented in a \emph{copula scatterplot} to obtain a useful visual tool to study dependencies. 

Given a $2$-dimensional sample of $d$ data $\{(X^{1}_1,X^{1}_2),(X^{2}_1,X^{2}_2), \dots, (X^{d}_1,X^{d}_2)\}$, we define a \emph{pseudo-observation} from the copula as 
\begin{equation}\label{eqn:pobs_2D}
\vec{U}^{i}_{} = (U^{i}_1,U^{i}_2) := \left(\hat{F}^{}_1(X^{i}_1), \hat{F}^{}_2(X^{i}_2)\right)
\end{equation}
for $i=1,2,\dots,d$. Here, $\hat{F}^{}_1$ and $\hat{F}^{}_2$ indicate the \acp{eCDF}:

\begin{equation}\label{eqn:ecdf_2D}
	\hat{F}^{}_1(x^{}_1) = \frac{1}{n} \sum_{i=1}^{d} 1^{}_{\{X^i_1 \le x_1\}} \qquad
	\hat{F}^{}_2(x^{}_2) = \frac{1}{n} \sum_{i=1}^{d} 1^{}_{\{X^i_2 \le x_2\}} \qquad
	x^{}_1,x^{}_2 \in \mathbb{R}
\end{equation}
\\

A scatterplot of $\vec{U}$, called \emph{copula scatterplot}, can be used as a graphic tool to illustrate dependencies between the involved \acp{RV}.\\


Extension of these ideas to higher dimensional copulas is immediate. 
However, when $n>3$ it is not feasible to visualize the copula scatterplot.

Summarizing, copula detaches the information contained in the joint behavior from the one due the marginals. Copula scatterplots illustrate the dependencies in the dataset from a graphical point of view, which, although being qualitative, encode a more global perspective compared to numeric coefficients like Kendall's $\tau$ or Spearman's $\rho$ \cite{nelsen2006introduction}. We remind here their empirical definition while we refer to \cite{nelsen2006introduction} for more details.
\\

\begin{description}
    \item[Kendall's tau]: Given a random sample $\{(x_1,y_1),(x_2,y_2),\dots,(x_n,y_n)\}$ of $n$ observation from the vector $(X,Y)$ of continuous \acp{RV},	let ($x_i,y_i$) and ($x_j,y_j$) denote two observations from ($X,Y$). We say that ($x_i,y_i$) and ($x_j,y_j$) are \emph{concordant} if
    $(x_i - x_j)(y_i - y_j) > 0$ and that they are \emph{discordant} in the opposite case. Let $c$ denote the number of concordant pairs and $d$ the number of discordant ones (so that $n=c+d$ is the total number of observation), the estimator of Kendall's tau for the sample, denoted as $\hat{\tau}$, is given by:
    \begin{equation}
    	\hat{\tau} := \dfrac{c-d}{n}
    \end{equation}
    \item[Spearman's rho]:
    When the  $n$ ranks of the observations are distinct integers, we can define the estimator of the Spearman's $\rho$ as: 
    \begin{equation}
    	\hat{\rho} := 1 - \frac{6 \sum n_i}{n(n^2-1)}
    \end{equation}
    Where $d_i = {R}(x_i) -{R}(y_i)$ is the difference between the two ranks of each observation.

\end{description}

 Independent \acp{RV}, characterized by the \emph{independent copula} \ref{def:ind_cop}, show a uniformly distributed scatterplot on $[0,1] ^2$. The presence of clusters of points or curves with higher densities reveals specific dependencies (see  \ref{sec:results}).
For a two dimensional copula the main diagonal (the one from the bottom left corner to the top right corner) of its scatterplot represents the points related by non-decreasing function $\phi$ such that $Y = \phi(X)$. Hence, these points correspond to a perfect (deterministic) correlation between those variables. So, the more the diagonal is densely populated, the more correlated are the variables.
Independent \acp{RV}, characterized by the \emph{independent copula}, show a uniformly distributed scatterplot on the unit square. 

\section{Copulas to detect dependencies}\label{sec:dependencies}
In this work we focus on different shapes of scatterplot associated to different types of dependencies.
Copulas capture dependencies between pairs of random variables. 
In \cite{sacerdote2012detecting} we applied copulas to synthetic neuronal data. There we considered two different scenarios. 
In the first one we assumed to have a sample of $n$ independent, identically distributed \acp{FPT} modelling the spiking times of a pair of neurons $(T_A^iT_B^i), i=1,..,n$ and we studied their copula. This instance corresponds to study the \acp{ISI} following synchronous spikes of the two neurons.  
In a second scenario, we called $S_A^i$ the spiking times of neuron $A$ and we  defined an interval $\theta^i$,  inter-time between the spike time of the neuron $A$  and the first spike of neuron $B$.   Then we considered a sample of pairs $(\theta_i, T_A^i)$, with $T_A^i=S_A^i-S_A^{i-1}$ and we studied their copula. In that work we also proposed to increase the understanding of the dependencies between spike trains exchanging the role of target neuron between $A$ and $B$.
Here, we aim at recognizing direct from indirect dependencies. Furthermore, we deal with networks of three neurons, with more complex dynamics. The use of the pairs of random variables proposed in \cite{sacerdote2012detecting} is not sufficient to disclose the structure of the links in the network. 
In particular, our interest on the direction of the links suggests to focus on the time flow when selecting the pairs of random variables for the copula. For example, using $A$ as target neuron  we expect independence between $T_A^i$ and $\theta^i$ if $B$ is pre-synaptic for $A$. However, in this case we expect to observe a dependence between $T_A^i$ and the time between the spike of $A$ and the former spike of $B$.  Hence, here we propose to add new pairs of \acp{RV} to our copula study.
%
To facilitate the reading, we introduce here all the pairs of \acp{RV} that we will use in Section \ref{sec:results}. \\
We first consider two neurons, denoted with $A$ and $B$. Following \cite{sacerdote2012detecting} We use 
$S^{}_A = \{S^1_A,\dots,S^n_A\}$ and $S^{}_B = \{S^1_B,\dots,S^m_B\}$ to indicate the spike times of $A$ and $B$ respectively, $S^i_C$ and $S^j_C$ being the epochs of the $i$-th and $j$-th events in the spike train $C=A, B$, for $i=1,\dots,n$ and $j=1,\dots,m$. Given a spike train, we define $i$-th \emph{forward} \ac{ISI}

\[
	T^{i}_{A,f} = S^{i+1}_A - S^{i}_A
	\]
	
	and $i$-th \text{backward}  \ac{ISI}:
	
	\[
	T^{i}_{A,b} = S^{i}_A - S^{i-1}_A.
	\]

	
An analogous convention is used for neuron $B$.

On a fixed recording interval, the number of spikes of two neurons might be different (i.e. $n \ne m$). To obtain a sample, we choose a \emph{target neuron}, for example $A$, and then define a way to pair each one of its \ac{ISI} with  significant intervals of the spike train of $B$. Again, we can proceed forward or backward (see Fig. 1):

\begin{enumerate}
	\item \textbf{Forward:} we define the interval $\Delta^i_{B,f}$ as the inter-time between $S^i_A$ and the first spike in $S^{}_B$ following it, denoted by $S^{i*}_{B,f} = \min \{\, S^i_B \in S_B | S^i_B > S^i_A\,\}$. 
	\[
	\Delta^i_{B,f} =S^{i*}_{B,f} - S^{i}_A.
	\]
	
	\item \textbf{Backward:} we define the interval $\Delta^i_{B,b}$ as the inter-time between $S^i_A$ and the first spike in $S^{}_B$ preceding it, denoted by $S^{i*}_{B,b} = \max \{\, S^i_B \in S_B | S^i_B < S^i_A\,\}$. 
	\[
	\Delta^i_{B,b} = S^{i}_A - S^{i*}_{B,b}
	\]
\end{enumerate}
In the forward case, the sample is determined by the pairs 
$\{(T^{1}_{A,f},\Delta^1_{B,f} )$, 
$\dots$, $ (T^{N}_{A,f},\Delta^N_{B,f} ) \}$, while in the backward case 
$\{(T^{1}_{A,b},\Delta^1_{B,b} ),$ $ \dots$, $(T^{M}_{A,b},\Delta^M_{B,b} ) \}$ is used. Here $N$ and $M$ are the sizes of the samples, that depend on $n$ and $m$.
Of course the role of $A$ and $B$ can be swapped, so that from a network of two neurons generating a pair of spike trains $S^{}_A$ and $S^{}_B$ it is possible to obtain four distinct samples:
\begin{enumerate}
	\item \textbf{{FWD - A}}: choosing $A$ as the target neuron in the forward approach.
	\item \textbf{{BWD - A}}: choosing $A$ as the target neuron in the backward approach.
	\item \textbf{FWD - B}: choosing $B$ as the target neuron in the forward approach.
	\item \textbf{BWD - B}: choosing $B$ as the target neuron in the backward approach.
\end{enumerate}

\begin{figure}
	\centering
	{\includegraphics[width = 0.5\textwidth]{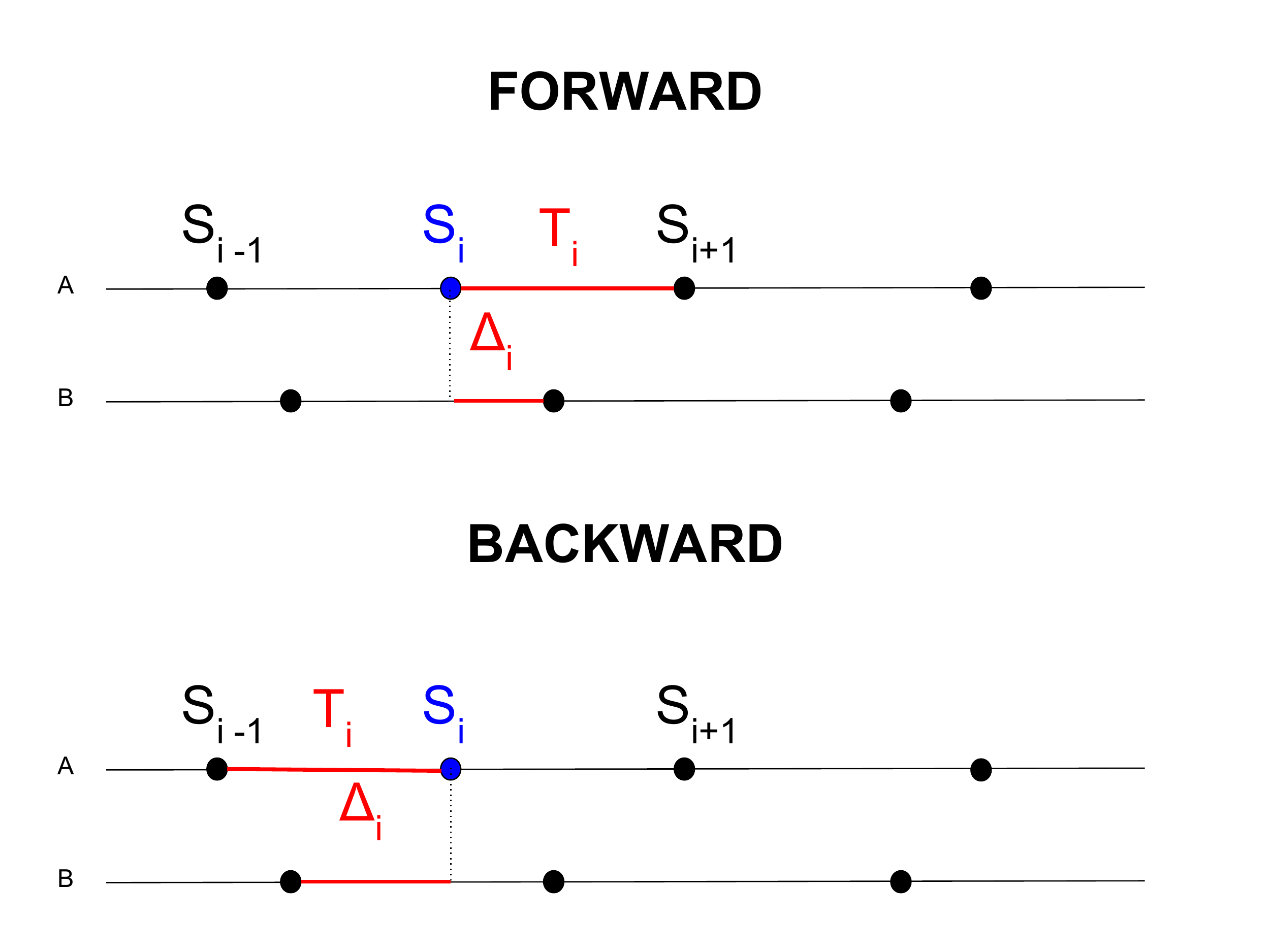} }
	\caption{Illustration of the procedure used to generate the pairs of intervals for the copula study using the forward (top) and the backward (bottom) approach, choosing $A$ as the target neuron.}
	\label{fig:STs}
\end{figure}

A graphical representation of this procedure is depicted in Fig.~\ref{fig:STs}. \\
In the following, when it does not generate misunderstanding, we use the notation $T^i$ to indicate a generic \ac{ISI} and $\Delta^i$ to referer to an inter-time, regardless of the fact that they refer $A$ or $B$ as a target neuron or that they were obtained by a forward or backward approach. When using copulas, we always indicate $U1$ (located in the $x$-axis) and $U2$ ($y$-axis) the \acp{CDF} of $T$ and $\Delta$, respectively.
\\
\\
A network of three neurons generates three spike trains and the number of combinations between forward/backward \acp{ISI} and backward/forward $\Delta$ increases.  Still, we proceed in an analogous way: denoting the neurons with $A$,$B$ and $C$ we choose one as a target (for example $A$ ) to obtain the \acp{ISI} ($T^{}_A$). Then, we compute the inter-times between the target and the other two neurons (namely, $\Delta^{}_{AB}$ and $\Delta{}_{AC}$). Since this can be done both choosing each one of the three neurons and proceeding backward or forward, we obtain six possible distinct cases. Those are:

\begin{enumerate}
	\item \textbf{{FWD - A}}: choosing $A$ as the target neuron in the forward approach.
	\item \textbf{{BWD - A}}: choosing $A$ as the target neuron in the backward approach.
	\item \textbf{FWD - B}: choosing $B$ as the target neuron in the forward approach.
	\item \textbf{BWD - B}: choosing $B$ as the target neuron in the backward approach.
	\item \textbf{FWD - C}: choosing $C$ as the target neuron in the forward approach.
	\item \textbf{BWD - C}: choosing $C$ as the target neuron in the backward approach.
\end{enumerate}

The joint analisys of copulas associated to these different intervals allows to guess the structure of the network. In Section 5 we illustrate how to work in this direction.

\section{Neural models for data generation}\label{sec:neural_models}

Neurons exhibit dependent spike trains when there exist links connecting them. However, dependent spikes arise also when two neurons receive their input form a common network.  Aim of this work is to show the usefulness of copulas to distinguish between different causes of observed dependencies. Here we extend to higher dimensions the two simplified models proposed in \cite{sacerdote2012detecting}. Using these models we produce synthetic data that we will use to apply the copulas method.

In both models we describe the \ac{MP} evolution using a multidimensional $\vec{X}(t) = \{ X_i(t) \}_{i =1}^N$. In this paper we consider $N = 2$ or $N=3$ and we assume that the membrane potential $X_i(t)$ of each neuron evolves as an \ac{OU} process \cite{sacerdote2013stochastic}:

\begin{equation} \label{eqn:OU}
\diff X_i(t) = \left(  -\frac{1}{\tau^{}_i} X_i(t) + \mu_i \right) \diff t + \sigma_i \diff W_i(t)
\end{equation}
Here $\tau_i \in \mathbb{R}_+$ and $\mu_i\in \mathbb{R}$,  $\sigma_i > 0$ are the \textit{decay time}, the {drift coefficient}  and  the {Noise intensity} of the $i-$th neuron, respectively. Furthermore  $W_i(t)$ is a standard \textit{Wiener process}, with null mean and unitary variance.

This model describes the evolution of the MP in the sub-threshold regime. When the depolarization reaches the threshold value $\theta$, the neuron elicits a spike and then it resets its membrane potential value to a resting value.
To introduce the dependency in the network of two or three neurons we propose to complete \eqref{eqn:OU} adding  further assumptions to mimic direct or indirect links between the neurons. Hence, we obtain two alternative models: the correlated Noise Model and the Jump model. 

\subsection{Indirect links: Correlated Noise Model} \label{subsec:cor_noise}

\begin{figure}[ht!]
    \centering
    \includegraphics[width= 1.\textwidth]{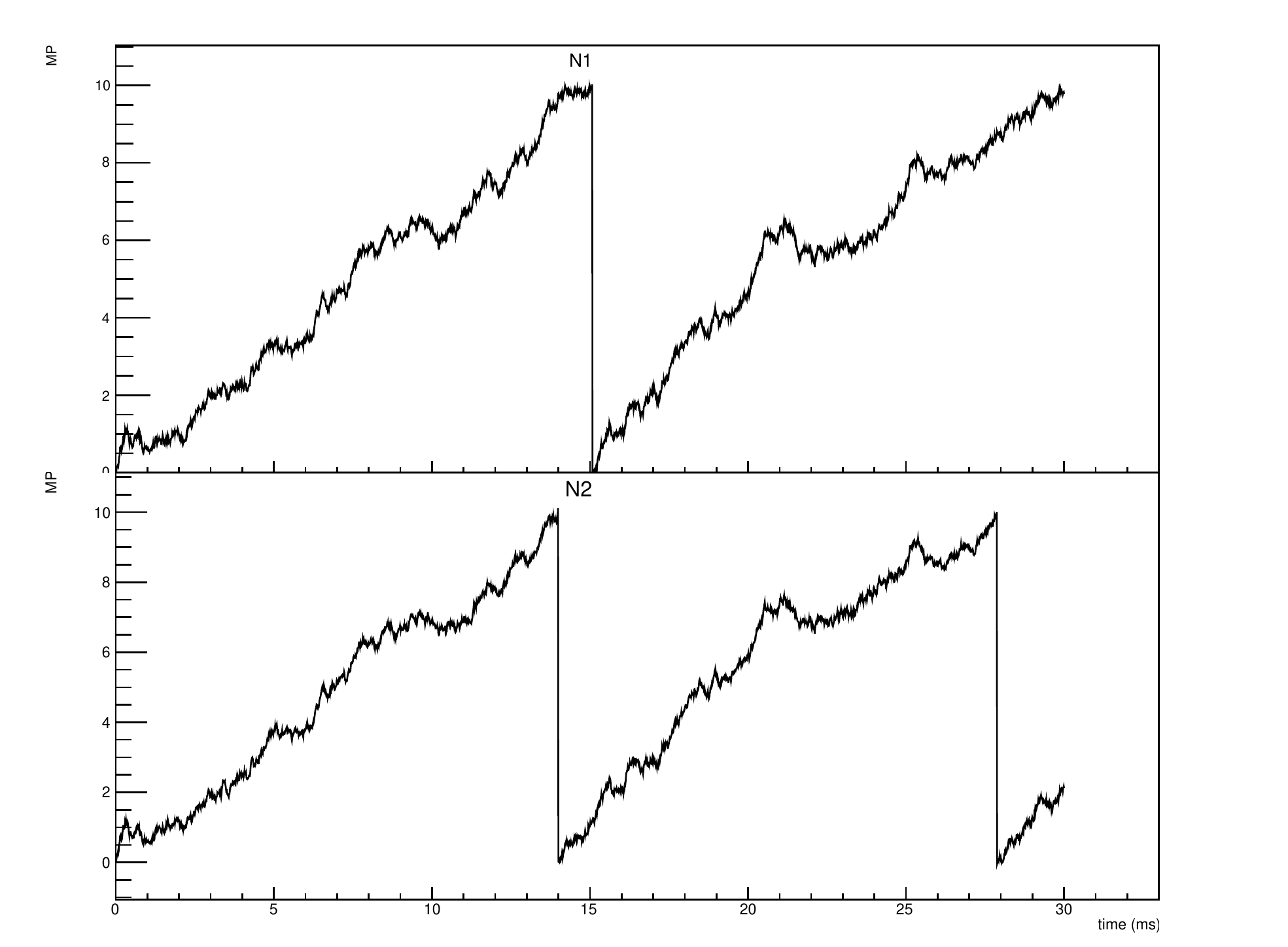}
    \caption{Example of the evolution of the \acp{MP} of two neurons using the correlated noise model ($c = 0.91$).}
    \label{fig:cov_process}
\end{figure}

In this model the Wiener processes $W_i$ in \eqref{eqn:OU} are not independent (see \cite{tamborrino2014weak}), but they satisfy $\text{Cov}[W_i,W_j] = \sigma_i \sigma_j c_{ij} $. The coefficients $c_{ij}$ may be positive, negative or null when $i \neq j$ while $c_{i,i}=1$.  In the first case (see Fig. \ref{fig:cov_process}), the membrane potentials of  neurons $i,j$ are positively correlated while the second case accounts for negative dependencies. Finally, null covariance implies the independence of the two neurons. The introduced dependencies among \acp{MP} induce dependent \acp{ISI}.

\subsection{Direct links: Jump Model} \label{subsec:jump}

\begin{figure}[ht!]
    \centering
    \includegraphics[width=1.\textwidth]{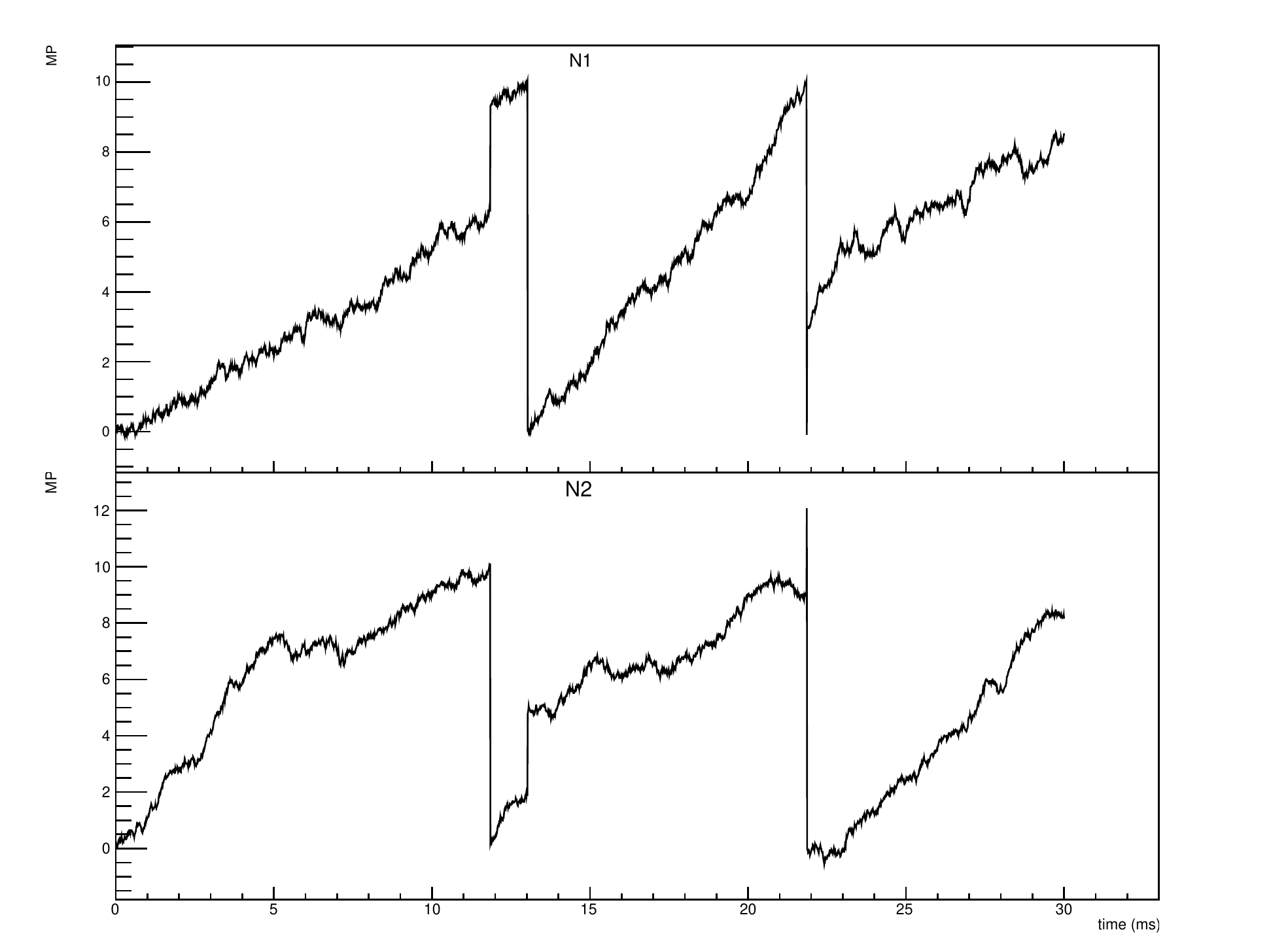}
        \caption{Example of the evolution of the \acp{MP} of two neurons using the jump model with positive (h=$3m$v) jumps.}
    \label{fig:jump_process}
\end{figure}

In this model each neuron \ac{MP} evolves independently from the others according to an \ac{OU} process \eqref{eqn:OU}, until the time when one of \ac{MP}s attains the threshold. Then this neuron elicits a spike and its \ac{MP} is reset to its resting potential while the \ac{MP}s of the other neurons of the network have an instantaneous variation, whose magnitude and nature (jump or drop) depends on the connection between the neurons. Different alternatives can be considered. For example,  each neuron exhibits a positive jump when the other attains the boundary (see Fig. \ref{fig:jump_process}) or one of two neurons has a negative jump or only one neuron has discontinuous sample paths.
\\

\section{Results}\label{sec:results}

We consider here a set of examples to illustrate the role  of the copulas to recognize links and dependencies. We generate synthetic spike trains using the models of Section \ref{sec:neural_models} and we distinguish two scenarios, following the schema used in \cite{sacerdote2012detecting}.
First, in Subsection \ref{subsec:FPT}, we reset the membrane potentials of the simulated neurons after the attainment of the threshold value, i.e. in the first scenario we study the \ac{FPT} of the depolarization process through a boundary. Despite the minor biological meaning this analysis is interesting because the initial conditions are the same for the involved neurons. Hence, it becomes easier to comprehend the nature of dependencies induced between neurons. Part of this study was already performed in \cite{sacerdote2012detecting}. 
In Subsection \ref{subsec:ST2} we discuss the analysis of spike trains, using the backward and forward times methodology described in \ref{sec:dependencies}. 
Then in Subsection \ref{subsec:ST3} we consider a network of three neurons and we show the use of forward and backward times to perform an analysis of the dependencies in the simulated spike trains.

The parameters used in the simulations are based on the values found in literature \cite{sacerdote2013stochastic,tuckwell1989stochastic,tuckwell1988introduction_2} for \ac{LIF} models. They are reported in Table~\ref{tab:std_val}. From now on we will refer to them as \emph{standard parameters} (or \emph{values}), mentioning explicitly every change, when present.
To generate  samples of spike trains, we simulated spike trains for a time interval $L^{}_t=250\,$s.  To remove a possible correlation due to the fact that neurons at $t=0$ share the same \ac{MP} value (as they both start from resting state), we extracted the first pair $(T^1, \Delta^1)$ (for all the four possible combinations) after the $50^{th}$ spike in the time series, discarding the previous ones. The samples dimension is $N=10^4$.

We do not perform any sensitivity analysis with respect to the standard parameters while we consider different values for the correlation $c$ of the \emph{Correlated Noise Model} and for the jumps size of the \emph{Jump Model}.

\begin{table}[th]
	\centering
	\caption{\small Standard values for the \emph{drift coefficient} $\mu$, the \emph{membrane constant} $\tau$, the (squared) \emph{noise intensity} $\sigma^2$ and the threshold $\tau$.}
	\label{tab:std_val}
	\begin{tabular}{c c c c }
		\toprule
		$\mu$			&$\tau$		&$\sigma^2$		&$\theta$\\
		(mV/ms)			&(ms)		&(mV$^2$/ms)	&(mV)\\
		\midrule
		1.2				&10			&0.3			&10\\
		\bottomrule
	\end{tabular}
	
\end{table}

\subsection{First Passage Times}\label{subsec:FPT}
We start with the case of two neurons and we study dependencies between \acp{FPT} induced by dependencies between membrane potential dynamics. We consider different values for the correlation coefficient $c$ of the \emph{Correlated Noise Model} of \ref{subsec:cor_noise}. The \acp{FPT} of the two neurons are dependent due to the presence of  correlated noise  in the membranes potential dynamics. We consider different values of $c$. In 
Table \ref{tab:corr}  we study the dependencies between the \acp{FPT} by means of the corresponding values of  Pearson's correlation coefficient, Kendall's tau and Sperman's $\rho$. The three coefficients have different sensitivity to changes of $c$. The values of Spearman's rho are the most close to those of $c$.

%

\begin{table}
	\centering
	\caption{\small Values of the estimated Pearson's correlation coefficient ($\hat{r}$), Kendall's tau ($\hat{\tau}$) and Spearman's rho ($\hat{\rho}$) for the correlation model, using different values  of $c$. All the values are statistically different from zero, since the p-values are smaller than $0.05$. }
	\begin{tabular}{S S S S}
		\toprule
		{$c$} 	& {$\hat{r}$} 	& {$\hat{\tau}$} 	& {$\hat{\rho}$} \\
		\midrule
		 0.5		& 0.38 			& 0.28 				&	0.40 \\
		 0.8		& 0.68			& 0.52  			&	0.68 \\
		 0.91		& 0.80			&0.68 				&	0.83\\
		 -0.91		&-0.56 			&-0.48 				&	-0.70\\
		\bottomrule
	\end{tabular}
	
	\label{tab:corr}
\end{table}


In the \emph{Jump Model}  different combinations of \acp{PSP} can be simulated, using different jump matrix. We call $h_{12}$ the value of the jump induced in the second neuron when the first one fires, while $h_{21}$ denotes the \emph{viceversa}.
Table \ref{tab:jump} reports the values for the Pearson's correlation coefficient, Kendall's tau and Spearman's $\rho$ between \acp{FPT} for different values of $h_{12}$ and $h_{21}$. Higher values of the jumps generate higher dependencies between \acp{FPT}, and this phenomenon is recognized by all the considered indices.

\begin{table}
	\centering
	\caption{\small Values of the estimated Pearson's correlation coefficient ($\hat{r}$), Kendall's tau ($\hat{\tau}$) and Spearman's rho ($\hat{\rho}$) for the jump model, using different values of $h_{12},h_{21}$.  All the values are statistically different from zero.  }
	\begin{tabular}{S S S S S}
		\toprule
		{$h^{}_{12}$} & {$h^{}_{21}$} 	& {$\hat{r}$} 	& {$\hat{\tau}$} 	& {$\hat{\rho}$} 	\\
		\midrule
		 1			&1					&0.34			&0.35				&0.43				\\
		 3			&3					&0.93			&0.92				&0.94				\\
		 3			&0					&0.49			&0.45				&0.55 				\\
		 3			&1					&0.62			&0.63				&0.70				\\
		 -1			&-1					&-0.23			&-0.18				&-0.30				\\
		 -3			&-3					&-0.47			&-0.29				&-0.53				\\
		 3			&-3					&0.35			&0.32				&0.36				\\
		\bottomrule
	\end{tabular}	
	\label{tab:jump}
\end{table}

\begin{figure}[ht]
	\centering
	\subfloat[][ $X$ excites $Y$  \label{fig:link_e} ]
	{\includegraphics[width = 0.4 \textwidth]{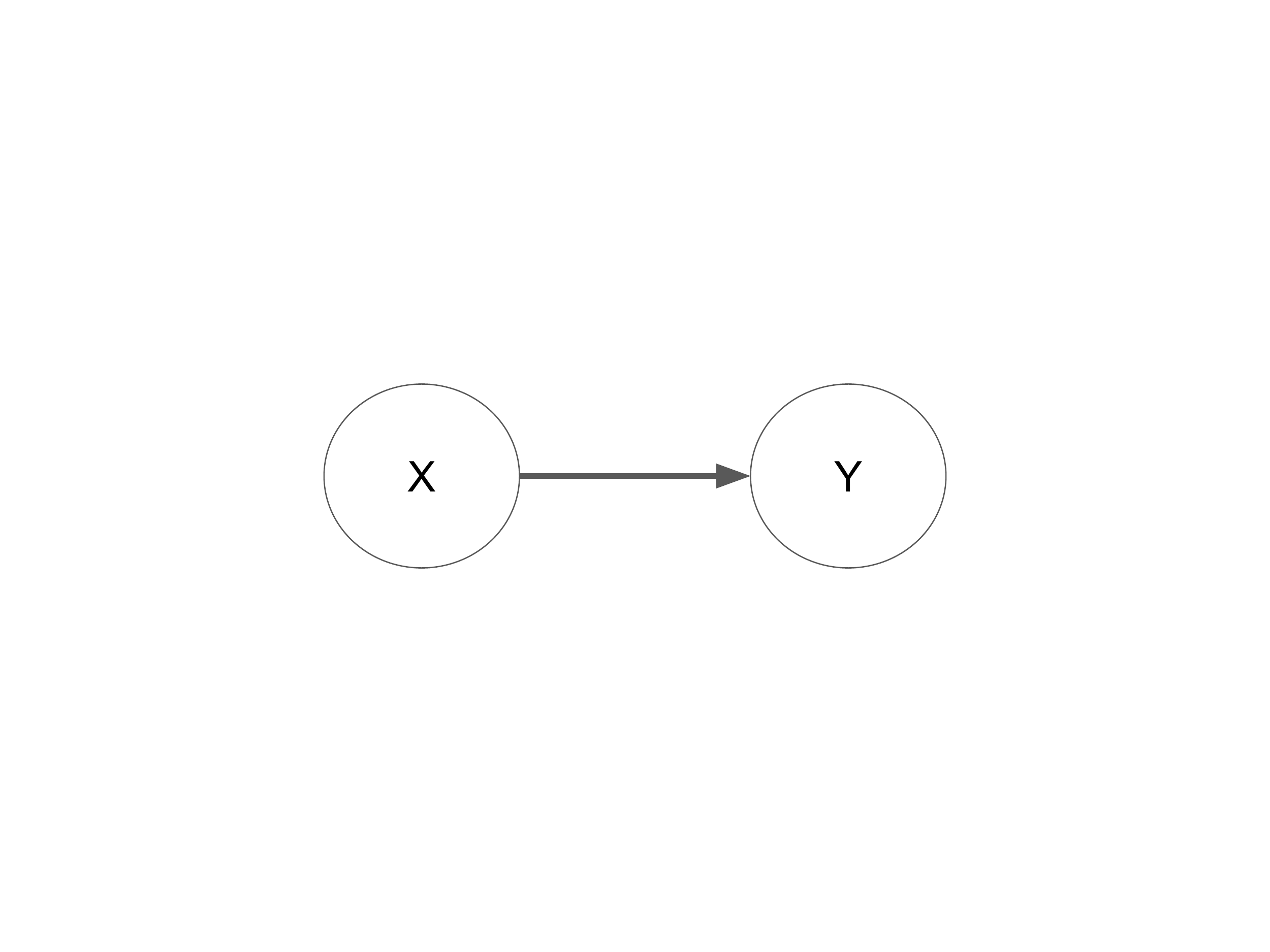}} \quad
	\subfloat[][ $X$ inhibits $Y$\label{fig:link_i}]
	{\includegraphics[width = 0.4 \textwidth]{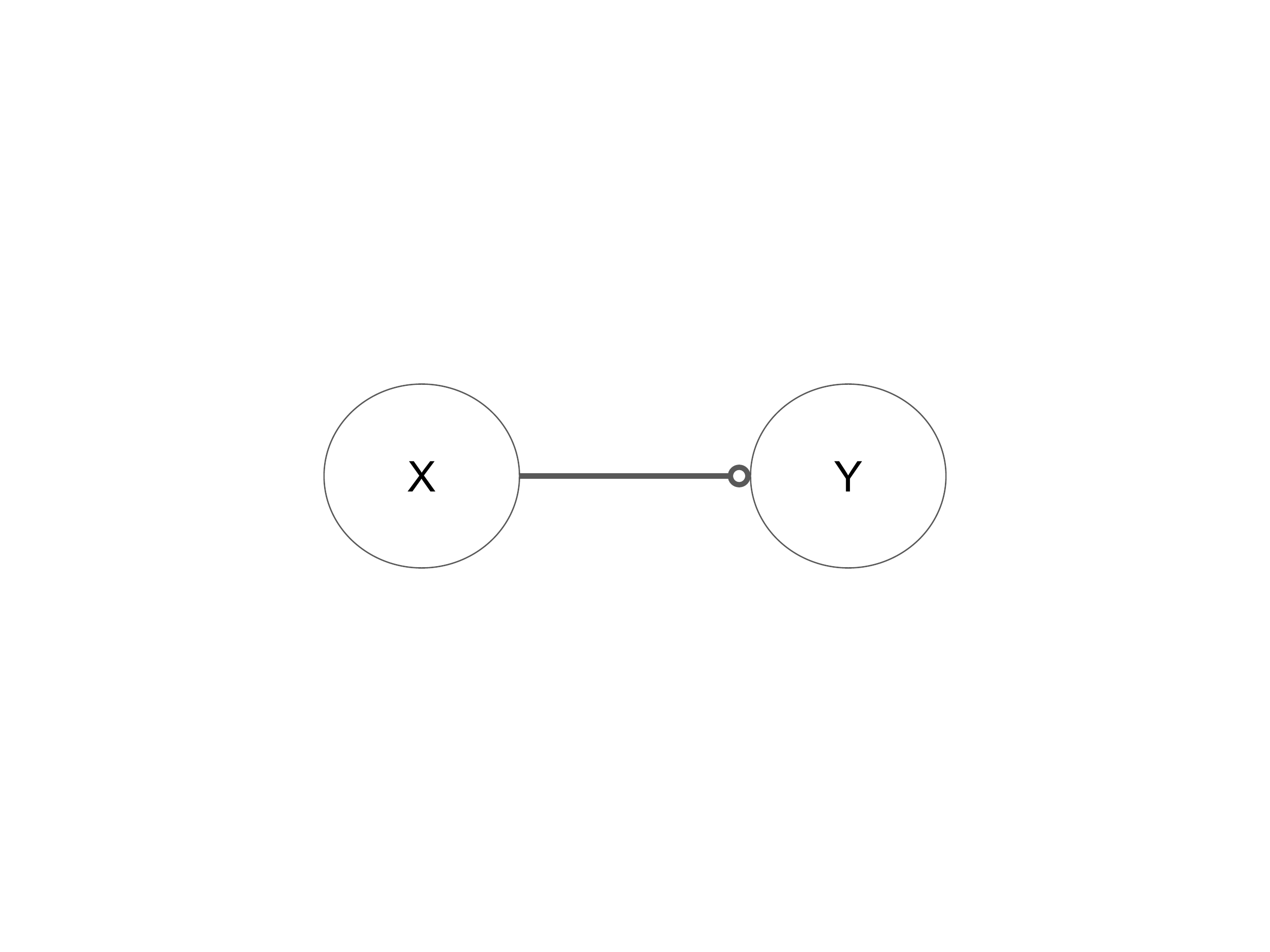}}
	\caption{Excitatory and inhibitory bonds.}
	\label{fig:link}
\end{figure}

We refer to \cite{sacerdote2012detecting} for examples of scatterplots of copulas of the \acp{FPT} corresponding to the \emph{Jump} or the \emph{Correlated Noise Models}. \\
Adding a third neuron does not change the copulas between spiking times of the \emph{Correlated Noise Model}. On the contrary, the situation for the \emph{Jump Model} becomes more complex since it increases the number of combinations of possible connections. To represent those links, we introduce a graphical visualization in which \emph{excitatory} connection are drawn as \emph{arrows} while \emph{inhibitory} connections are drawn as \emph{circles} (see Fig.~\ref{fig:link}). \\
We illustrate how to get an intuition on the nature of connections by means of some examples focusing on \acp{FPT} of three neurons. We generally use the same magnitude of $3$mV for the jumps, positive or negative to avoid to introduce further variability among experiments, simplifying the comparison. 
Exceptions to this choice will be reported.

Three distinct bi-dimensional copulas are associated to the \acp{FPT} of this network, each one encoding the dependency between different pairs of neurons. In the following, we report figures illustrating dependencies between the three neurons of the network. Each figure presents 9 sub-panels illustrating the dependencies between pairs of neurons. The 3 sub-panel on the diagonal are clearly empty, while those above the diagonal are the same as those below but with the axes exchanged.

\begin{figure}
	\centering
	\subfloat[][
	\label{fig:E38}] 
	{\includegraphics[width = 0.33\textwidth]{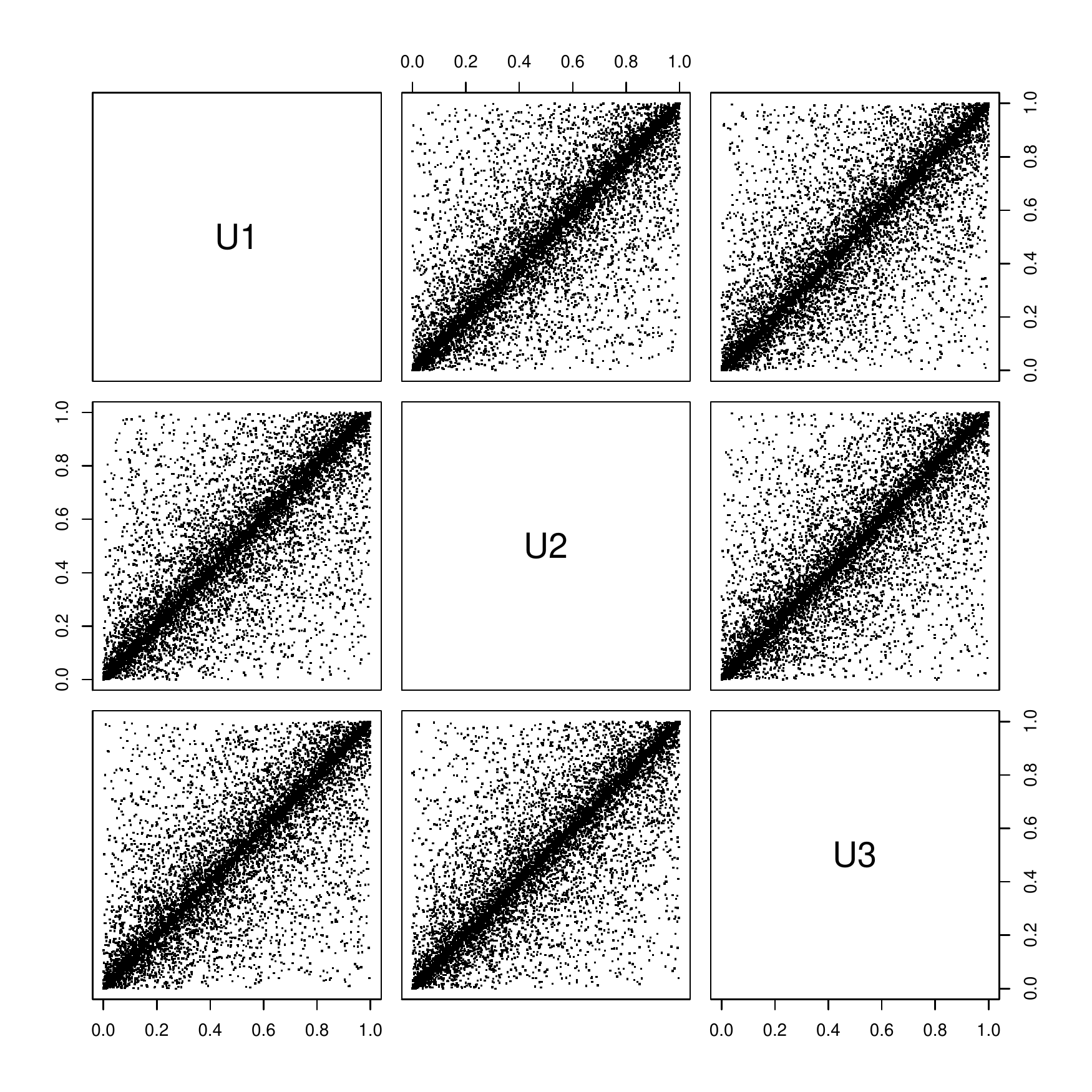}}
	\subfloat[][
	\label{fig:E40}
	] 
	{\includegraphics[width = 0.33\textwidth]{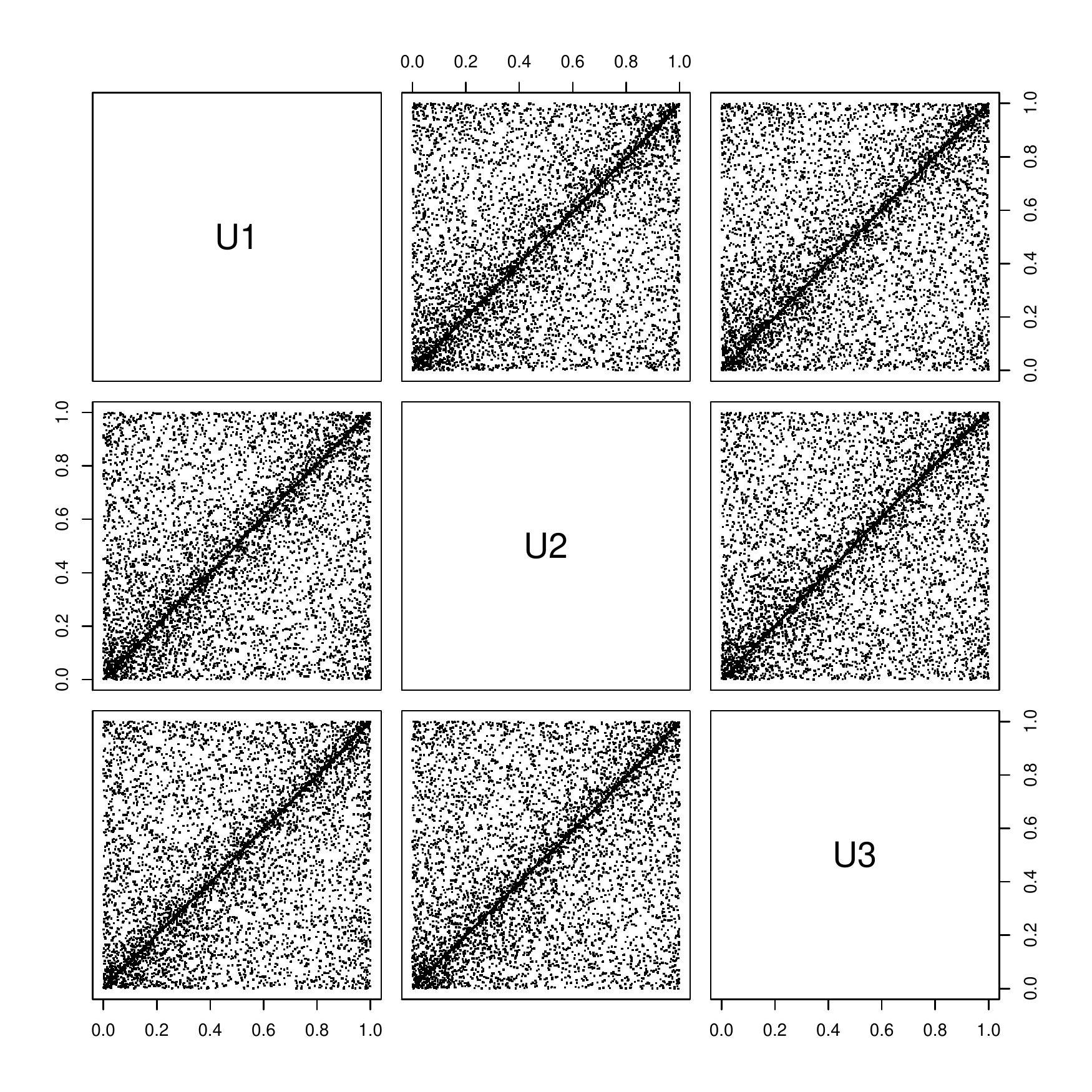}}
	\subfloat[][
	\label{fig:E14_g}]
	{\includegraphics[width = 0.3\textwidth]{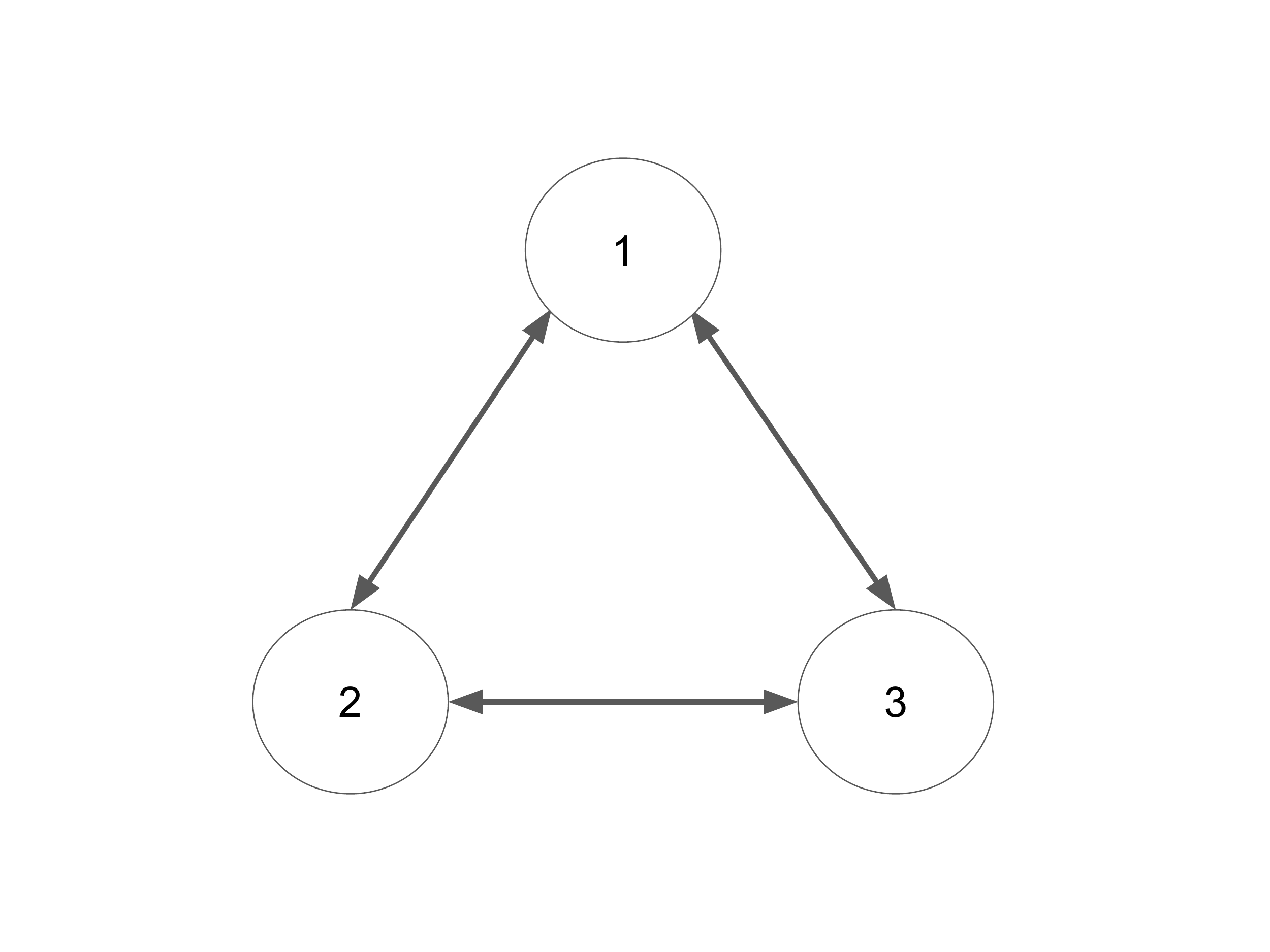}}
	\caption {
	Copula scatterplots for the  \acp{FPT}. Panel \protect\subref{fig:E38} shows the $4$ copulas obtained using the correlated noise model with $c=0.91$, while panel~ \protect\subref{fig:E40} displays the copulas obtained using the jump model, with the network represented in  \protect\subref{fig:E14_g}. Here all jumps have the same height $h = 1$mV. Note that the values of the Kendall's tau (Tab.\ref{tab:taus_3D} are similar
	}
	\label{fig:E1432}
\end{figure}

Let us observe Fig.~\ref{fig:E38} and \ref{fig:E40} ignoring the procedure used to generate the scatterplots. 
In the scatterplots of both panels of the figure we observe many points on the diagonal. They reveal a strong correlation between the \acp{FPT} of involved neurons.   However, while the clean lines of all the scatterplots in panel~\subref{fig:E40} suggest the existence of direct links between neurons, the presence of many dispersed points around the diagonal in the scatterplots of panel~\subref{fig:E38}  is compatible with the presence of a common noise influencing all the three neurons. Furthermore, since scatterplots corresponding to neurons (1)-(2), (1)-3 and (2)-(3) in Fig.~\ref{fig:E40} are identical we conclude that the cause of the observed dependencies should be the same for the three involved neurons. Hence, we expect a network like the one in panel~\ref{fig:E14_g}. These intuitions are confirmed revealing the model generating the analyzed synthetic data. 


Figure \ref{fig:E38} is generated using the correlated noise model ($c = 0.8$).There  a dense cloud of points concentrated around the diagonal substitutes  the clear line of synchronicity. This observation allows us to clearly distinguish this case from data generated using the \emph{Jump Model}. Hence,  copulas allow to recognize the existence of direct links between neurons from the case of dependencies determined by a common noise. \\

\begin{figure}
	\centering
	\subfloat[][
	\label{fig:E17}] 
	{\includegraphics[width = 0.45\textwidth]{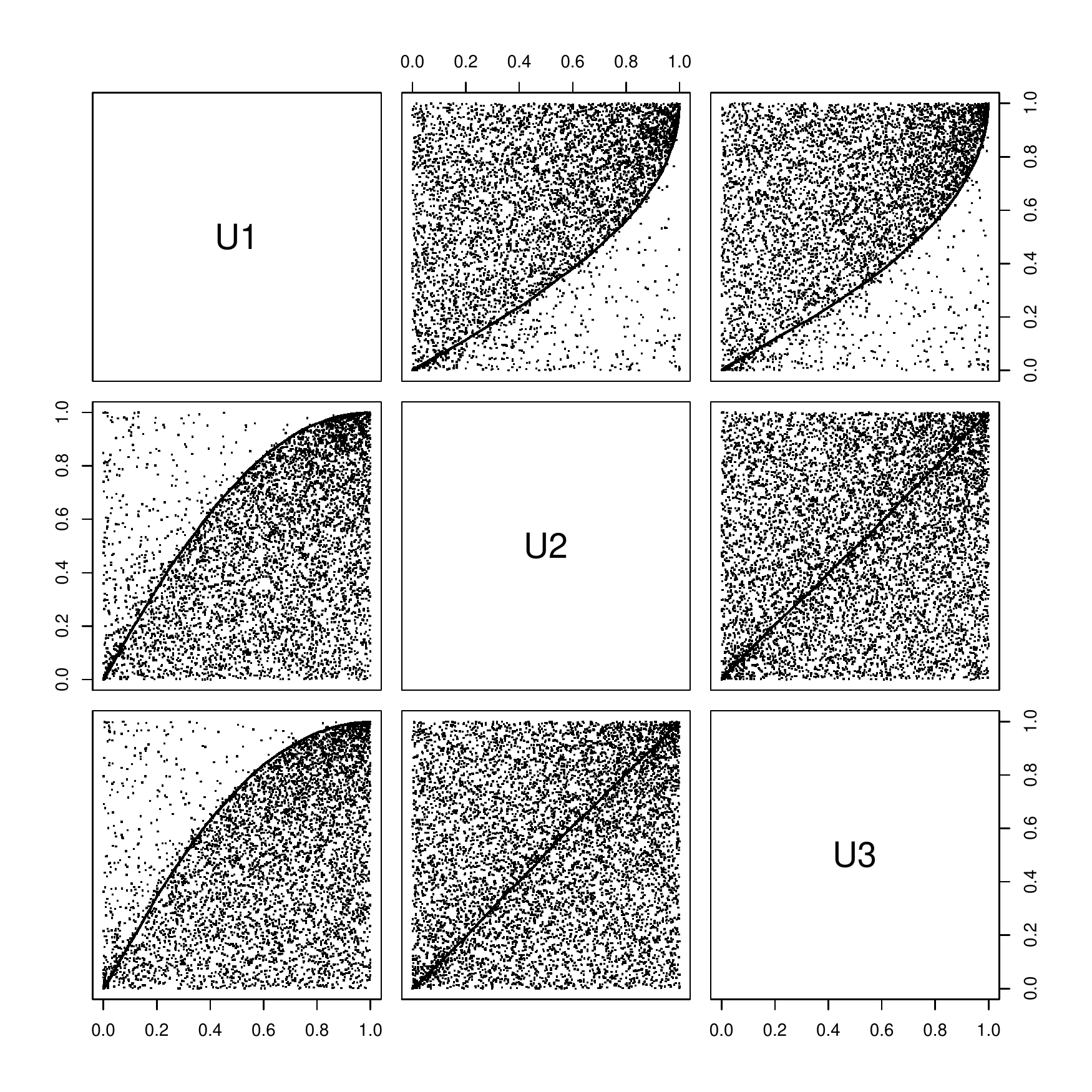}}
	\subfloat[][
	\label{fig:E32}] 
	{\includegraphics[width = 0.45\textwidth]{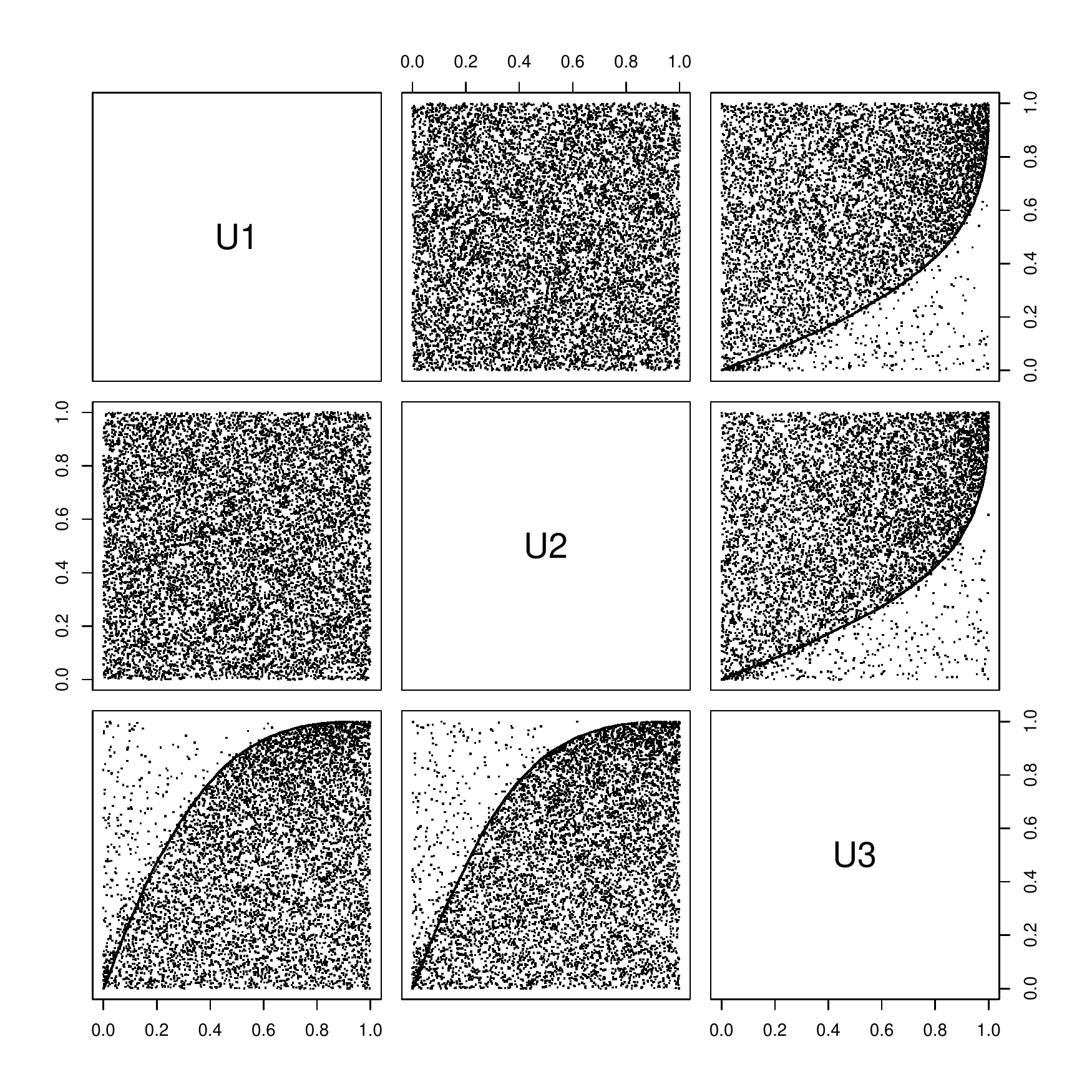}}	
	\\
	\subfloat[][
    \label{fig:E17_g}] 
	{\includegraphics[width = 0.45\textwidth]{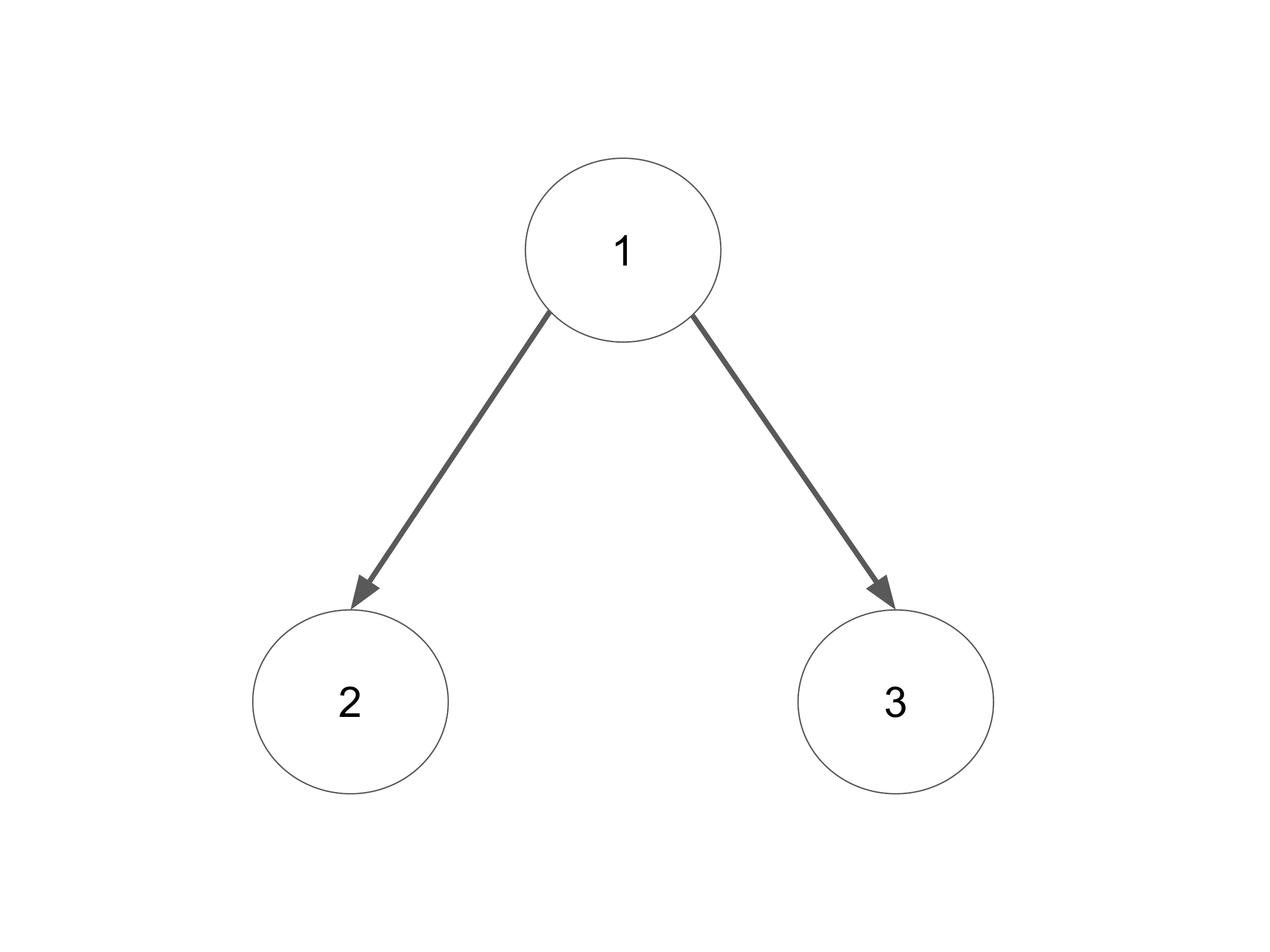}}
    \subfloat[][
    \label{fig:E32_g}] 
	{\includegraphics[width = 0.45\textwidth]{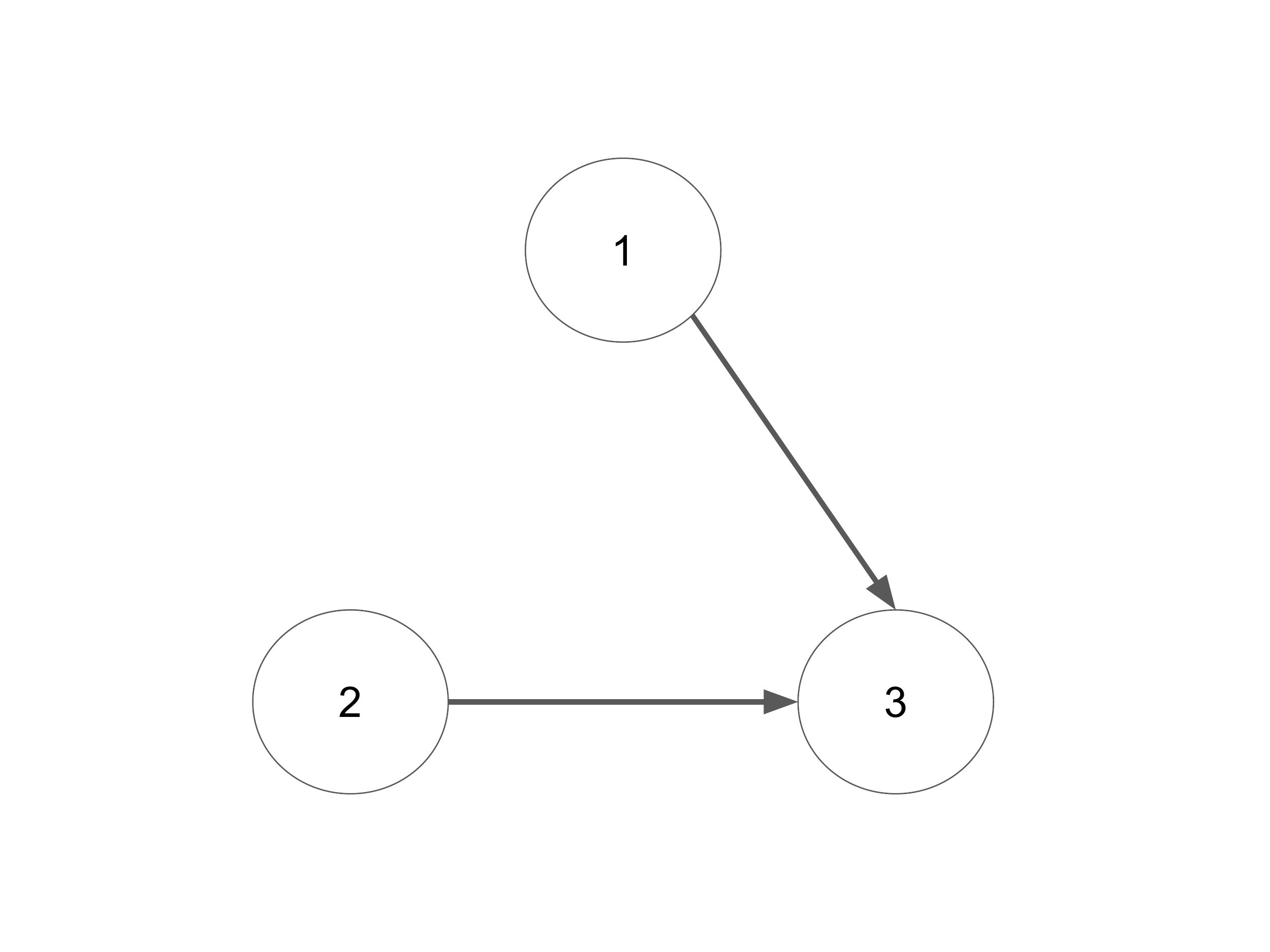}}
	\caption {Copula scatterplots for the \acp{FPT}. Panel~\protect\subref{fig:E17} is generated using the network represented in~\protect\subref{fig:E17_g} while panel~\protect\subref{fig:E32} is generated by the network in~\protect\subref{fig:E32_g}. Note how the synchronicity between neuron $(2)$ and $(3)$ in~\protect\subref{fig:E17} and the independence of $(1)$ and $(2)$ in~\protect\subref{fig:E32} can be clearly identified. Here the arrows represent excitatory connections with intensity $h=3$mV. }
	\label{fig:3217}
\end{figure}

Now, we focus on the possibility to guess the structure of the network from the scatterplots of \acp{FPT}.
%
Let us now consider Fig.~\ref{fig:E17}. The scatterplots of copulas between neurons $1-2$ and $1-3$ are very similar and are asymmetric. On the contrary, the copula between neurons $2-3$ shows a populated diagonal, indicator of synchronicity. In this last  case we also observe dispersed points, a fact reflected in a low tau value. 
Here, the points do not form a cloud around the diagonal. 
Hence, we cannot interpret these points  as in  Fig.~\ref{fig:E38}. However, these points are compatible with the presence of an input from neuron (1) that determines the spike in one of the two neurons (2) or (3) but not necessarily in both.  The observed asymmetry of the other scatterplots  may correlate to the common influence of neuron (1) on both neurons (2) and (3), facilitating their spiking activity. A minor number of spikes results related to the spontaneous activity of each neuron.
Values of dependency indices are coherent with this hypothesis. In fact this experiment almost have a perfect correlation, as indicated by the values of Kendall's $\tau$ in \ref{tab:jump}.
These scatterplots are compatible with the case in which two neurons are driven by third one as shown in panel~\subref{fig:E17_g} of the Figure.

In Fig.~\ref{fig:E32} we immediately recognize the independent copula between neuron (1) and neuron (2).  This remark is confirmed by the null value of Kendall's tau. The two neurons do not exchange any information through common inputs.  Scatterplots between neurons $(1)-(3)$ and $(2)-(3)$ are similar and their shape reminds the shapes observed between neurons $(1)-(2)$ or $(1)-(3)$ in~\ref{fig:E38}. The two scatterplots displaying mono-directional bonds suggest the network in panel~\subref{fig:E32_g}, where neurons $(1)$ and $(2)$ only excite neuron $(3)$ without being excited. This agrees with the fact that those neurons do not exchange any information. The intuitions on the network structure correspond to the network used for the \acp{FPT} generation.

\begin{figure}
	\centering
	\subfloat[][
	\label{fig:E15}] 
	{\includegraphics[width = 0.45\textwidth]{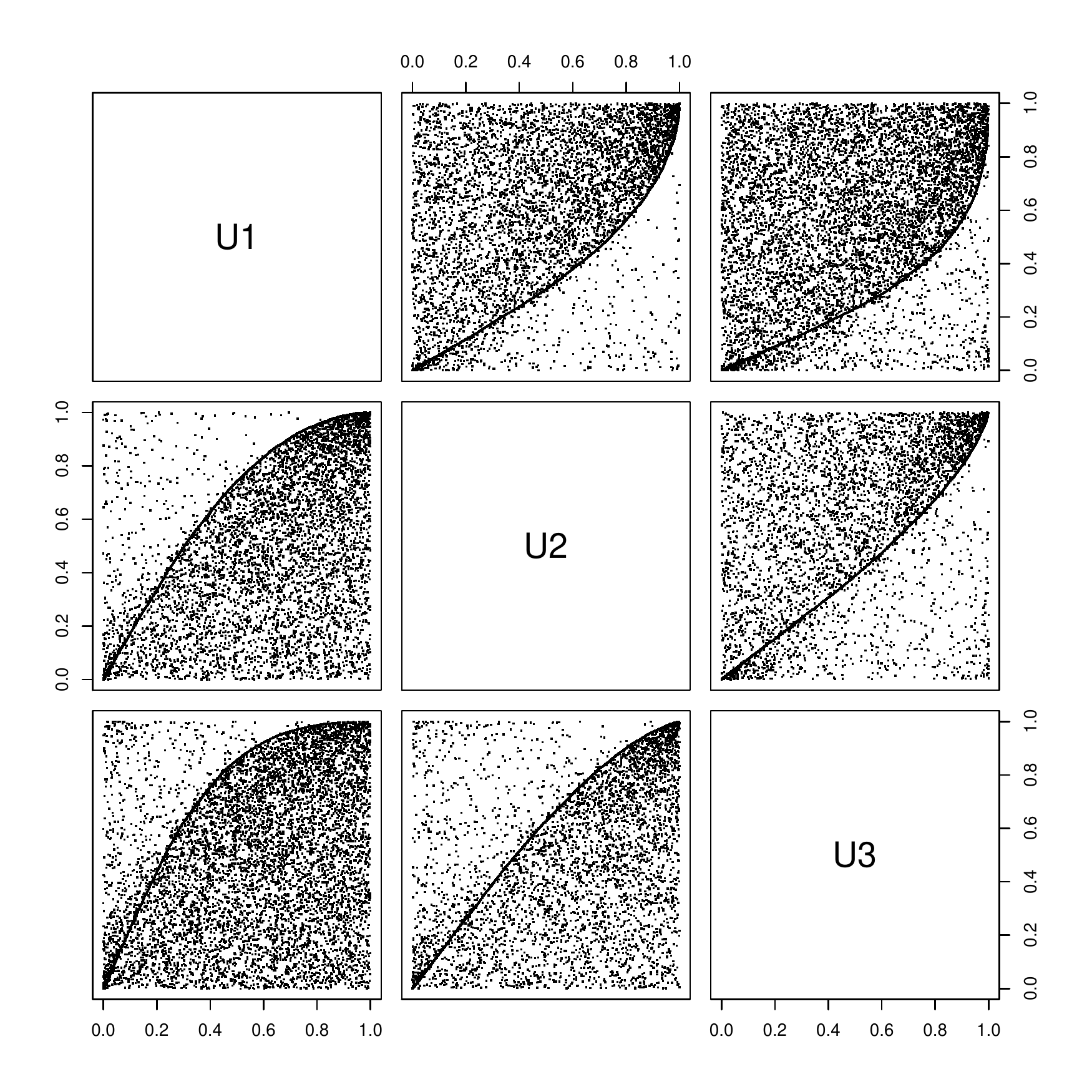}}
	\subfloat[][
	\label{fig:E33}] 
	{\includegraphics[width = 0.45\textwidth]{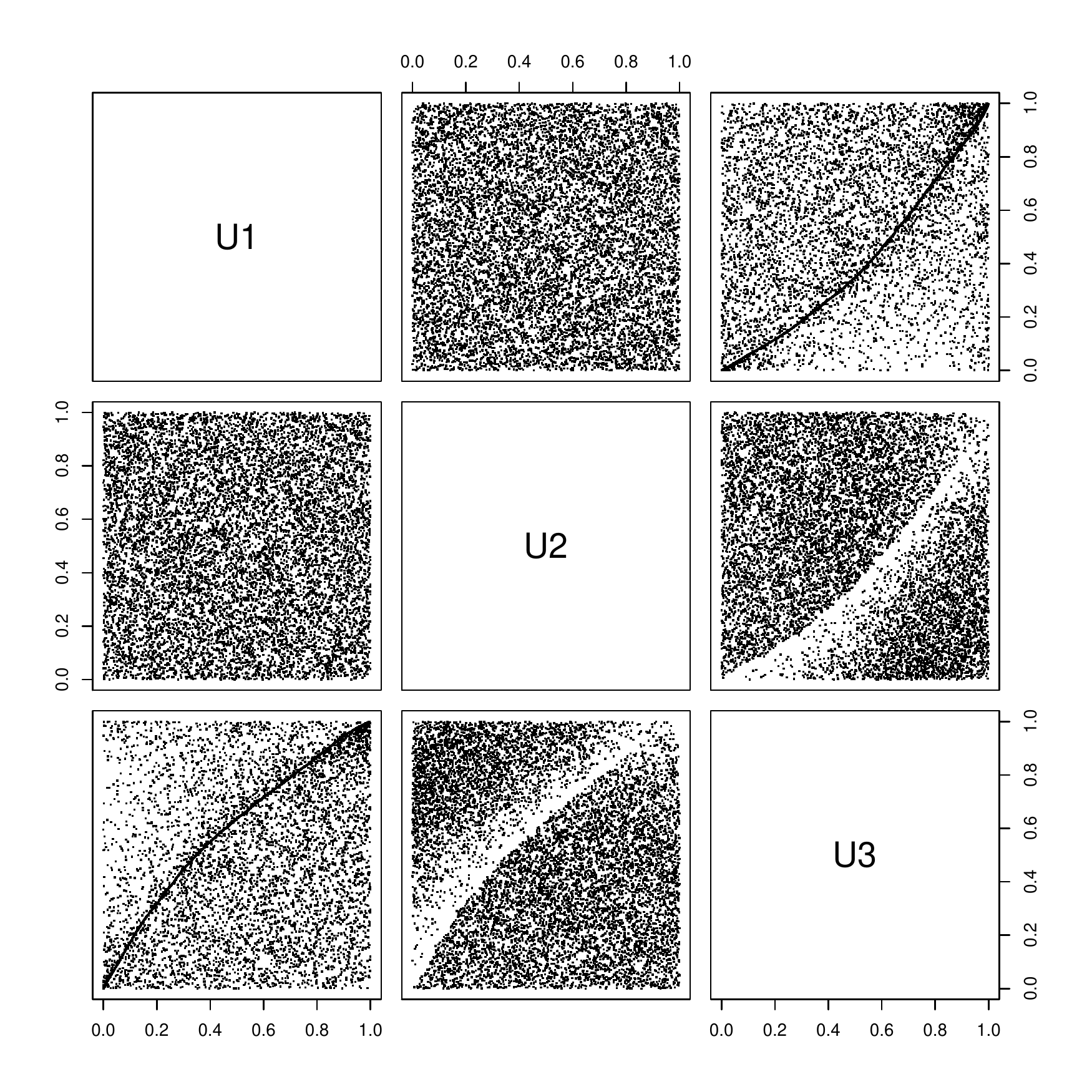}}
	\\
	\subfloat[][\label{fig:E15_g}]
	{\includegraphics[width = 0.45\textwidth]{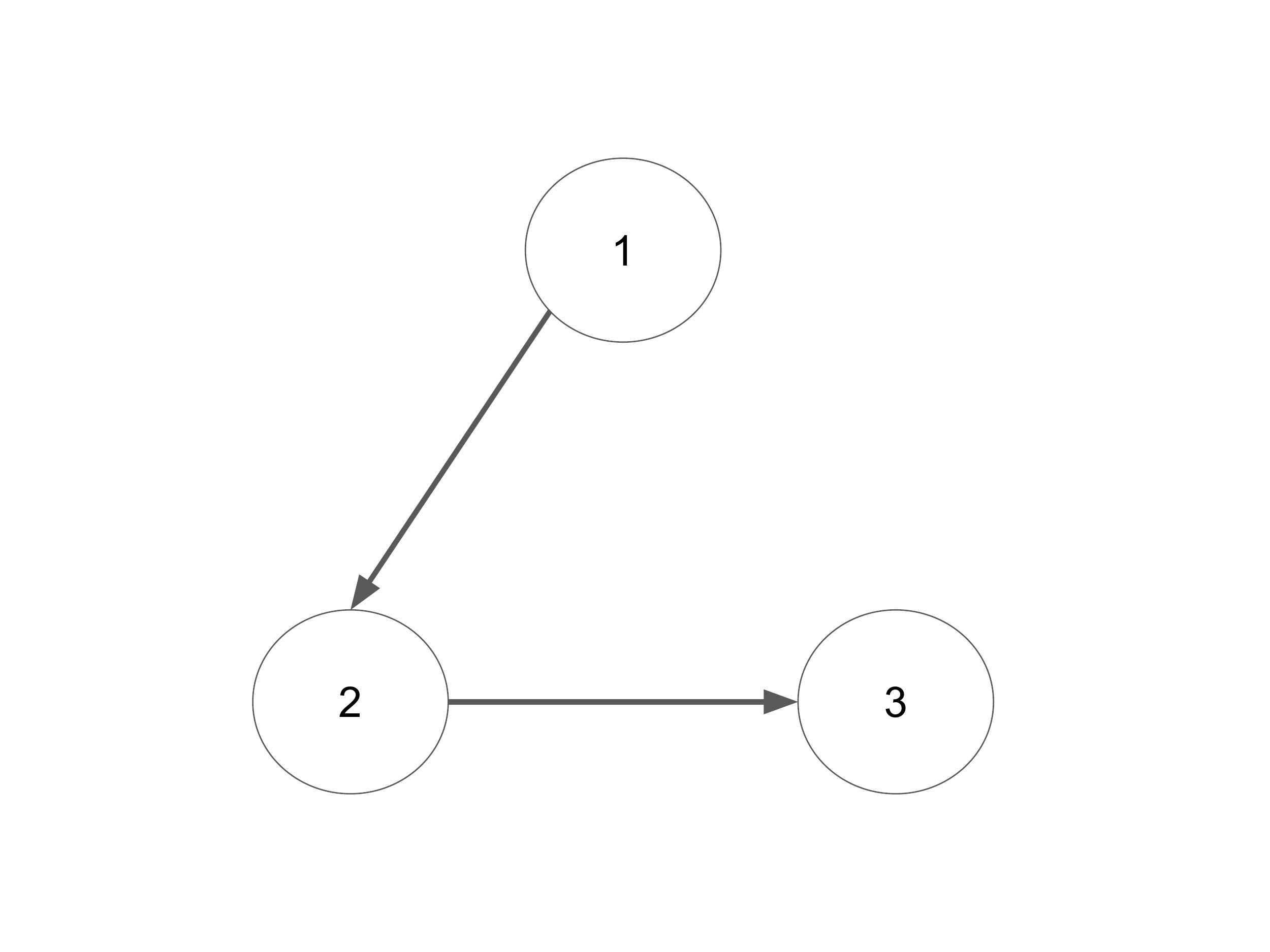}}
    \subfloat[][
    \label{fig:E33_g}]
	{\includegraphics[width = 0.45\textwidth]{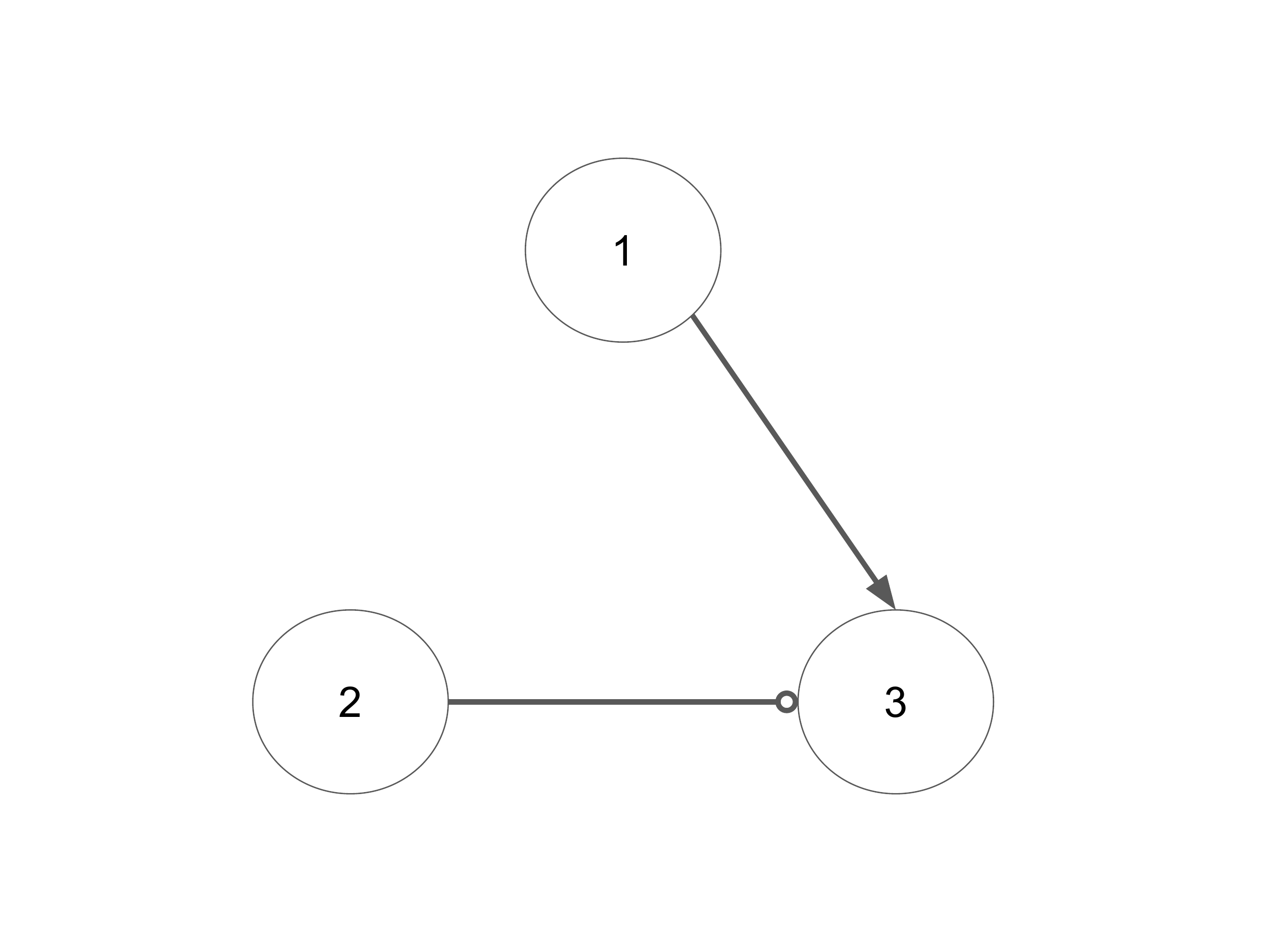}}
    \caption {Copula scatterplots fot the \acp{FPT}. Panel~\protect\subref{fig:E15} is generated using the network represented in~\protect\subref{fig:E15_g} while panel ~\protect\subref{fig:E33} is generated by the network in~\protect\subref{fig:E33_g}. Note how, in ~\protect\subref{fig:E15} the weaker bond between $(1)$ and $(3)$, which are not directly connected, is highlighted by a less-curved  synchronicity line. In ~\protect\subref{fig:E33} the absence of points under the synchronicity curve is the typical feature of inhibitory connection. Here the arrows represent excitatory connections with intensity $h=3$mV, while the inhibitory connection (represented by a circle) has a intensity of $-3$mV.}
		\label{fig:E1533}
\end{figure}

In Fig.~\ref{fig:E15} we recognize the dependency between firing times of neurons $(1)-(2)$ and  $(1)-(3)$, respectively. We note that their scatterplots are similar but, the synchronicity curves are ordered with that of neuron $(3)$ higher than for neuron $(2)$. The largest part of the points is below the synchronicity curve. It seems that the spike of neuron $(1)$ facilitates the spiking of neuron $(2)$ that eventually induces a spike of neuron $(3)$. This fact suggests sequential links of neuron $(1)$ with neuron $(2)$ and from  $(2)$ to $(3)$.

In Fig. ~\ref{fig:E33} the scatterplot between \acp{FPT} of neurons $(1)$ and $(3)$ is similar to those in Fig.~\ref{fig:E15} and we relate this result to the existence of excitatory inputs from $(1)$ and $(3)$. The scatterlots between neurons $(1)$ and $(3)$ show a new feature: the absence of points around the synchronicity curve. Note a highly non-trivial feature: the orientations of the curve (which are the same) indicates that the \ac{FPT} of neuron $(3)$ is more likely to be shorter than longer compare to the \acp{FPT} of neuron $(1)$ and $(2)$, respectively. This might be due to the fact that neuron $(3)$ can fire first (i.e. before both the others) a fraction of the times, while if it not first the effects of inhibition and excitation conflicts. A natural interpretation of this property suggests the existence of an inhibitory role of neuron $(2)$ on neuron $(3)$. Hence, we conjecture an excitatory effect of neuron $(1)$ on neuron $(3)$ and an inhibitory effect of this last neuron on neuron $(2)$. The presence of the independent copula between neurons $(1)$ and $(2)$ confirms this conjecture. Panels ~\subref{fig:E33_g} and ~\subref{fig:E15_g} show the graphs of the networks used to simulate data, confirming the intuition coming from the scatterplots.

\begin{table}
	\centering
	\caption{Values of the Kendall's tau for the $3$-dimensional \acp{FPT}. We use $0$ to indicate a measure with a p-value $>0.05$.  }
	\begin{tabular}{c  S S S}
		\toprule
		Simulation			&{$\hat{\tau}^{}_{12}$} & {$\hat{\tau}^{}_{23}$} 	& {$\hat{\tau}^{}_{13}$} \\
		\midrule
		\ref{fig:E38} 	&0.67	&0.67	&0.67 \\
		\ref{fig:E40} 	&0.48	&0.48	&0.48 \\
		\ref{fig:E17} 	&0.28	&0.46	&0.28 \\
		\ref{fig:E32} 	&0	&0.30	&0.31 \\
		\ref{fig:E15} 	&0.47	&0.28	&0.53 \\
		\ref{fig:E33} 	&0	&0.44	&0.14 \\
		\bottomrule
	\end{tabular}	
	\label{tab:taus_3D}
\end{table}

\subsection{Spike Trains in the Two-Neuron case}\label{subsec:ST2}
The analysis of \acp{FPT} was facilitated by the use of the same random variables on both the axes of scatterplots. Wishing to analyze scatterplots of spike trains we use \acp{ISI} coupled with forward or backward times (described in Table~\ref{sec:dependencies}). Here the analysis is more complex because the different RVs on the two axes determine an asymmetry. Hence, we start from networks of two neurons to collect ideas that will be used for networks of higher dimension.

\begin{figure} 
	\centering
	\includegraphics[width=0.45 \textwidth]{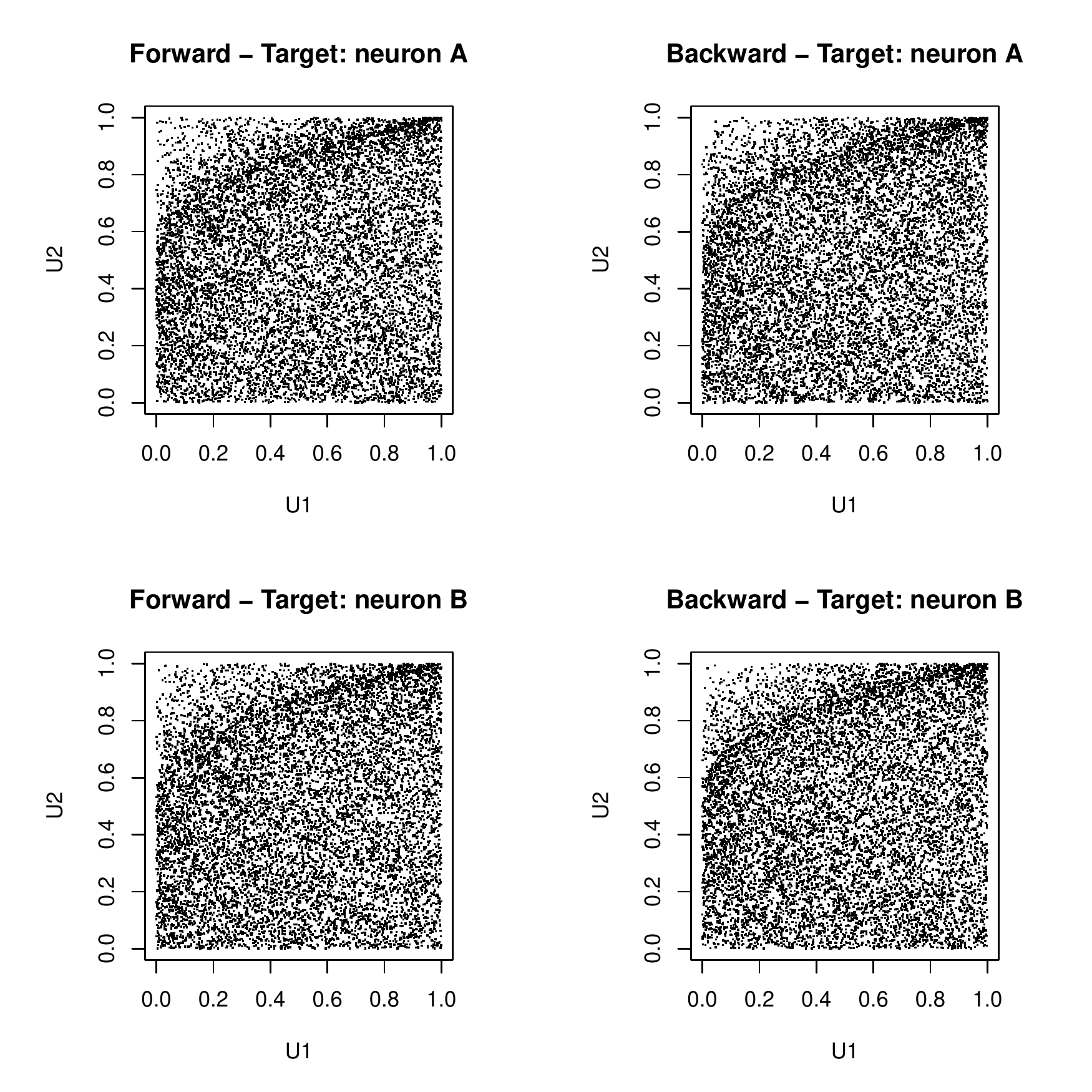}
	\caption{Copula scatterplots for the data generated with the correlated noise model, with $c=0.5$. See \ref{tab:E23} for the values of the coefficients.}
	\label{fig:E23}
\end{figure}

Let us first consider Fig.~\ref{fig:E23}.  All the four panels  appear  similar and so do all the values of the dependence coefficients reported in Table~\ref{tab:E23}. Since $T$ is an \ac{ISI} and $\Delta$ is the closest (backward or forward) spike of the other neuron it is  likely that $\Delta$ is shorter compared to $T$. However, this feature has no role on the copula since copulas are scale free. Furthermore,  the \acp{MP} do not start each time from the same (resting) value making the plots more noisy than those of \acp{FPT}. Hence we observe a weaker dependency in the dataset and we get smaller correlation coefficients for the same value of $c$ as in Fig.~\ref{fig:E40} (compare Table~\ref{tab:E23} with the first row of Table~\ref{tab:corr}). Fig.~\ref{fig:E23} reveals two aspects about neural dynamics of this experiment: a \emph{symmetry} in the dependency generation (since choosing $A$ and $B$ as a target does not make any significant difference) and an \emph{absence of a causal relation} between the neurons (since the backward and the forward approaches always looks the same). Hence, it is reasonable to attribute the observed dependence to a common phenomenon simultaneously influencing both neurons.
This analysis is in agreement with the model we used to simulate the data, since a correlated noise involves both neurons in the same way and do not introduce causality, as the two neurons do not communicate in any way. 

\begin{figure} 
	\centering
	\includegraphics[width=0.45 \textwidth]{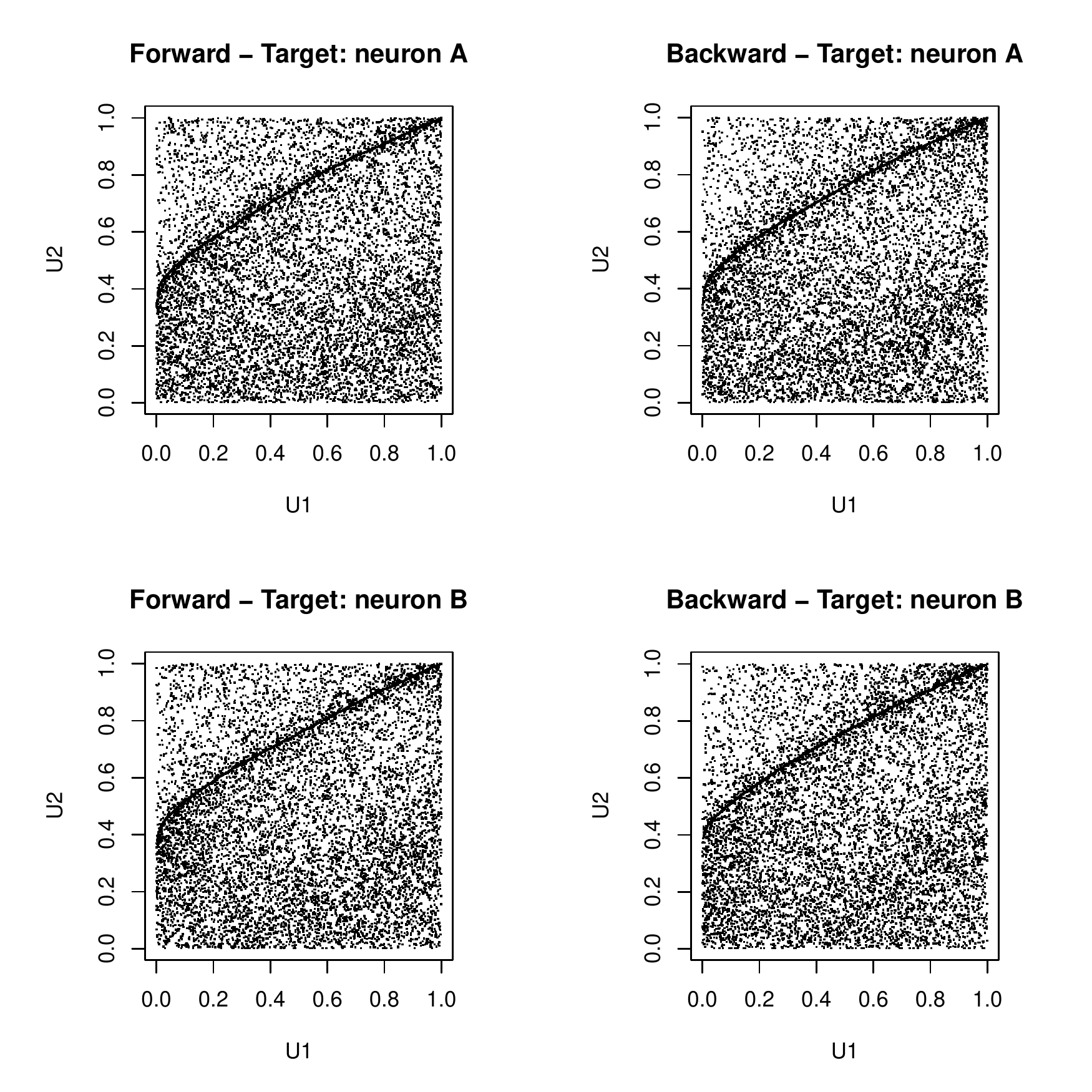}
	\caption{Copula scatterplots for the data generated with the jump model, $h^{}_{12}=1$ and $h^{}_{21}=1$. See \ref{tab:E36} for the values of the coefficients.}
	\label{fig:E36}
\end{figure}

In Fig.~\ref{fig:E36} we observe again a densely populated area but now it is delimited by a clean curve.  The well outlined line of synchronicity suggests the existence of direct links between the neurons. All the panels of the figure are very similar, suggesting a symmetry between the neurons, the spiking of one neuron influences the other and vice-versa. Furthermore, this figure  reminds us the scatterplot between neurons $2-1$ in Fig.~\ref{fig:E17} but this time it is more noisy.  A possible explication for this feature is the weakness of direct links between neurons. This feature is confirmed by the model used to generate synthetic data. Here $h_{12}=h_{21}=1$mV. Repeating the analysis with higher values of $h_{12}=h_{21}$ (not shown) we observe a decrease of the noise effect in the figure. Comparing with the \acp{FPT} that use the same jump intensities we note that, as usual, dependency is weaker in the data from the spike trains (see Tables \ref{tab:E36} and \ref{tab:jump}).


\begin{table} 
	\centering
	\caption{ Values of the estimated Pearson's correlation coefficient ($\hat{r}$), Kendall's tau ($\hat{\tau}$) and Spearman's rho ($\hat{\rho}$) for data generated using the correlated noise model, with $c=0.5$ (Fig.~\ref{fig:E23}) .
	All the values are statistically different from zero. }
	\begin{tabular}{c  S S S}
		\toprule
		Case 			  	& {$\hat{r}$} 	& {$\hat{\tau}$} 	&	{$\hat{\rho}$} 	\\
		\midrule
		FWD - A		& 0.16 	& 0.09 	& 0.13 \\
		BWD - A		& 0.19 & 0.11 & 0.16 \\
		FWD - B		& 0.17 & 0.10 & 0.15\\
		BWD - B		& 0.16 & 0.09 & 0.13  \\
		
		\bottomrule
	\end{tabular}	
	\label{tab:E23}
\end{table}

\begin{table}
	\centering
	\caption{ Values of the estimated Pearson's correlation coefficient ($\hat{r}$), Kendall's tau ($\hat{\tau}$) and Spearman's rho ($\hat{\rho}$) for data generated using the jump model, with $h^{}_{12}=1$ and $h^{}_{21}=1$ (Fig.~\ref{fig:E36}).
		All the values are statistically different from zero. }
	\begin{tabular}{c  S S S}
		\toprule
		Case 			  	& {$\hat{r}$} 	& {$\hat{\tau}$} 	&	{$\hat{\rho}$} 	\\
		\midrule
	
	FWD - A	& 0.15 & 0.13 & 0.16\\
	BWD - A		& 0.19 & 0.15 & 0.19\\
	FWD - B	& 0.16 & 0.14 & 0.17\\
	BWD - B		& 0.20 & 0.16 & 0.21\\
		\bottomrule
	\end{tabular}	
	\label{tab:E36}
\end{table}

\subsection{Spike trains in the Three-Neuron case}\label{subsec:ST3}
To make more visible the effect of links between neurons, in this paragraph the jumps always have a constant intensity of $3$mV.
In the scatterplots a capital letter indicates use of \acp{ISI}  for that neuron, while we use \texttt{D\textunderscore K} to indicate that we are considering the inter-times (backward of forward) between $K$ and target neurons .

\begin{figure}
	\centering
	\subfloat[][\label{fig:E28_A_FWD}] 
	{\includegraphics[width = 0.5\textwidth]{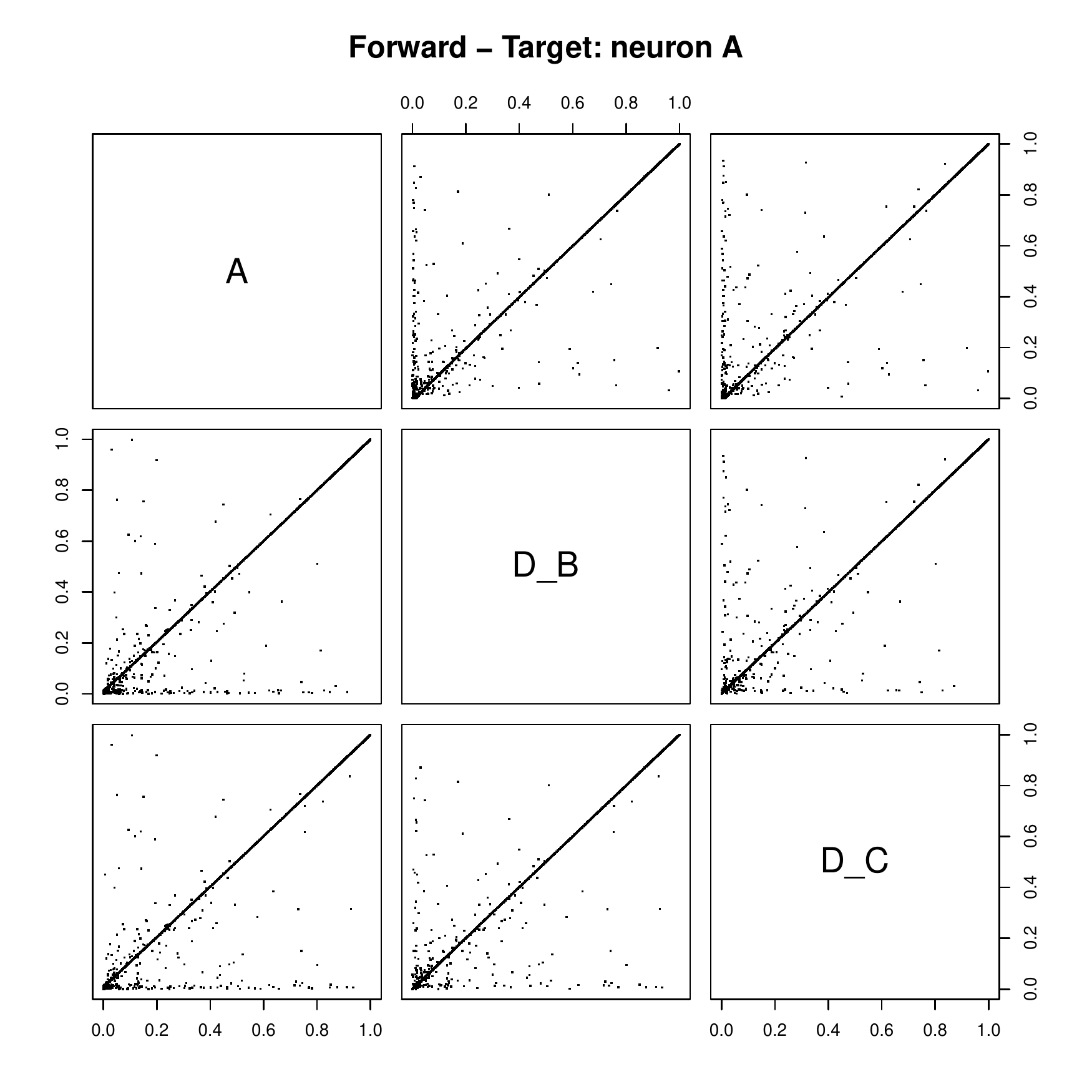}}
	\subfloat[][\label{fig:E28_A_FWD_g}]
	{\includegraphics[width = 0.5\textwidth]{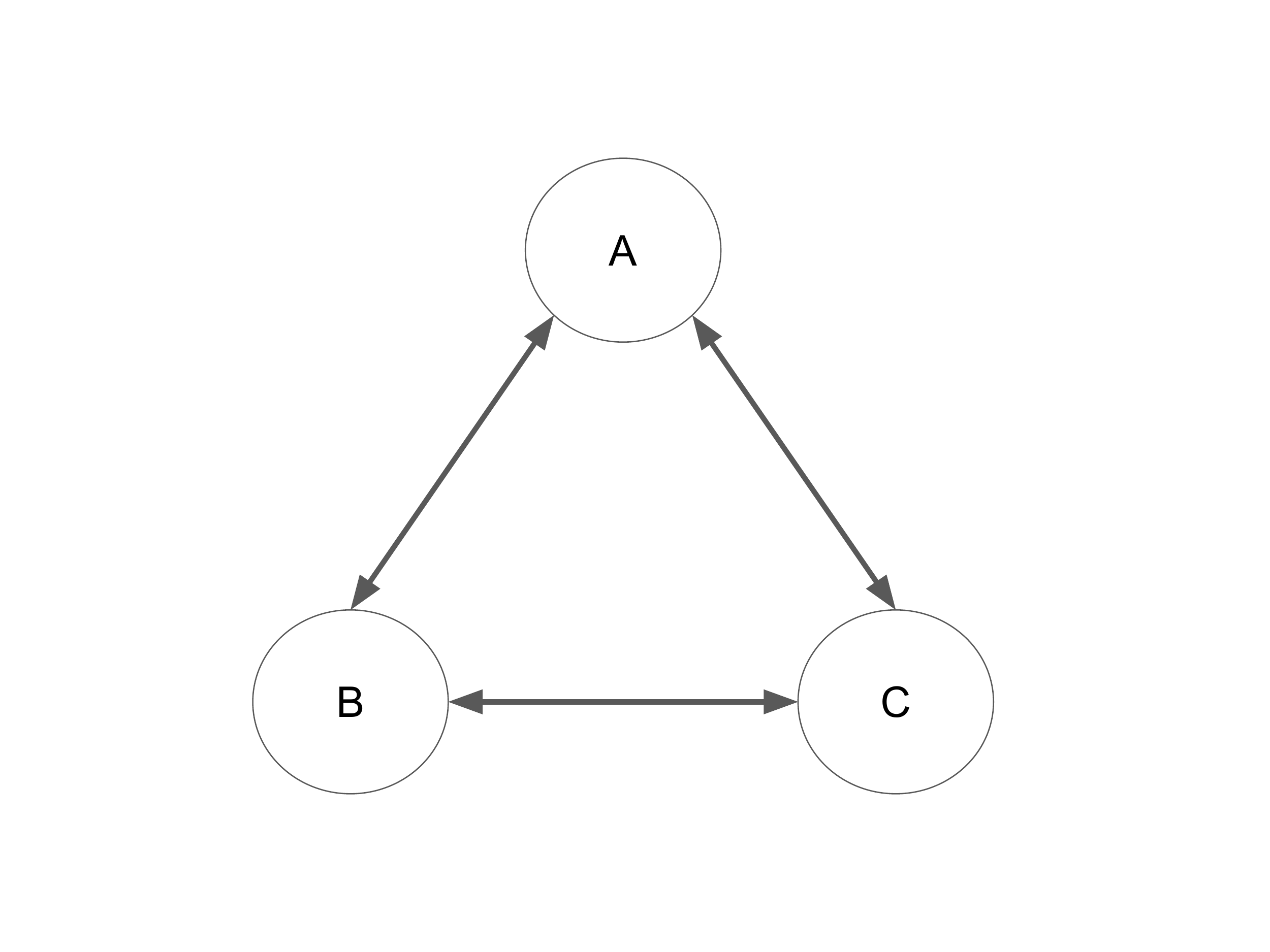}}
	\caption {Scatterplots for the spike trains and neural network used to generate them. Note the almost total synchronicity due to the fully-connected structure of the underlying network. Here the arrows represent excitatory connections with intensity $h=3$mV.\\ }
	\label{fig:E28}
\end{figure}

Fig.~\ref{fig:E28} is very similar to  Fig.~\ref{fig:E40}, that we recognize as its \ac{FPT} analogous. The main difference is the presence of clusters of points in the proximity of axes, for small values of \texttt{D\textunderscore B} and \texttt{D\textunderscore C}. This feature can be related with the random initial value of \acp{MP} generating the spike trains. Note that in Fig.~\ref{fig:E28}, first row and first column panels analyze the joint behaviour of an \ac{ISI} of a neuron and a forward time of another. This is in agreement with the analysis performed in the case of two neurons. However, the other panels of the figure compare \texttt{D\textunderscore B} and \texttt{D\textunderscore C}, two inter-times. We recognize straight and clearly-marked lines of synchronicity typical of  direct connections. Here, we  only show the group of scatterplots corresponding to $A$ target neuron  and forward times but all other alternative groups look similar to Fig.~\ref{fig:E28} (figures not shown). This result suggests the presence of  strong reciprocal relations and we conclude that the data are generated by a fully connected network, as shown in \ref{fig:E28_A_FWD_g} and confirmed by the knowledge of the  model used to generate the data.

\begin{figure}
	\centering
	\subfloat[][\label{fig:E29_A_FWD}] 
	{\includegraphics[width = 0.3\textwidth]{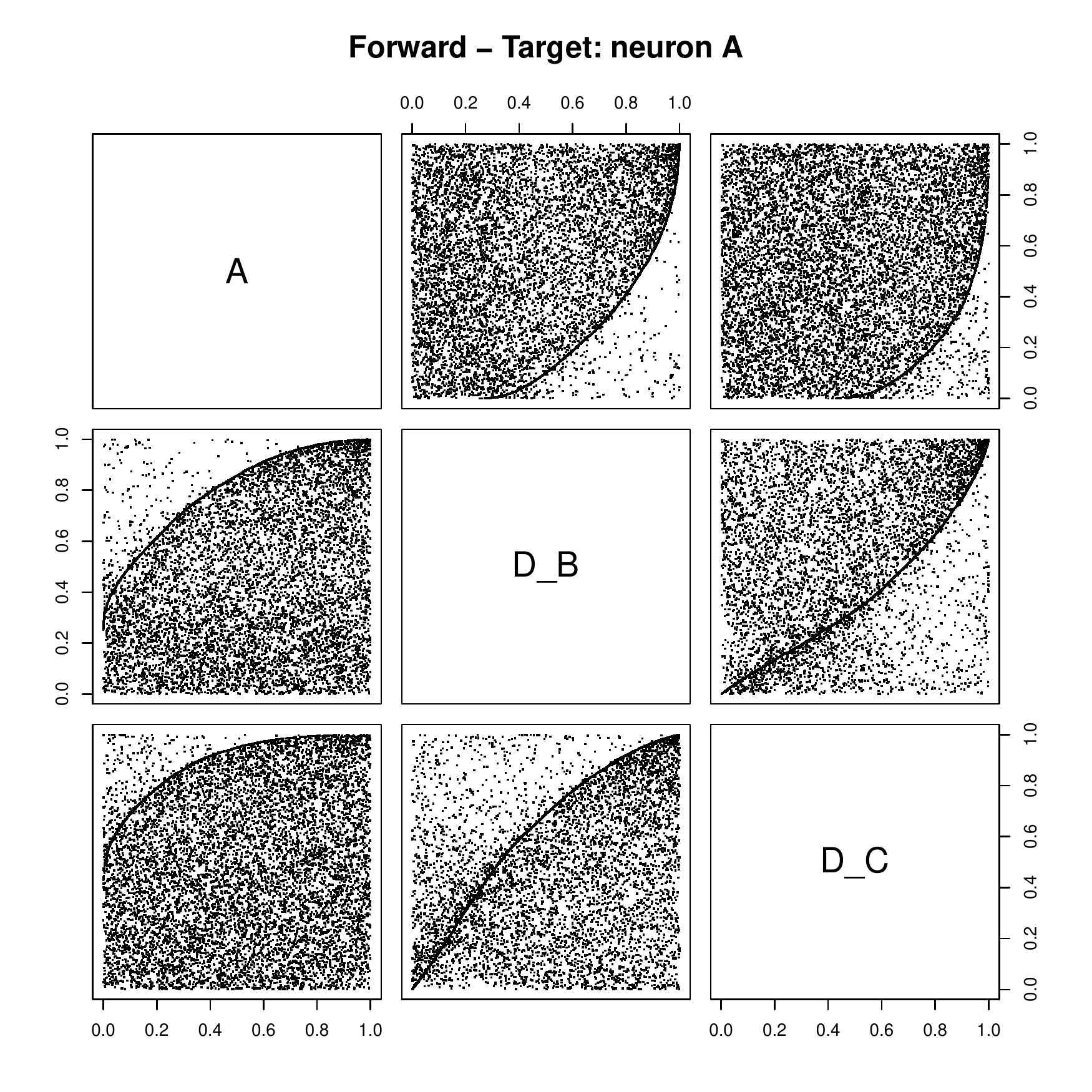}}
	\quad
	\subfloat[][\label{fig:E29_C_BWD}] 
	{\includegraphics[width = 0.3\textwidth]{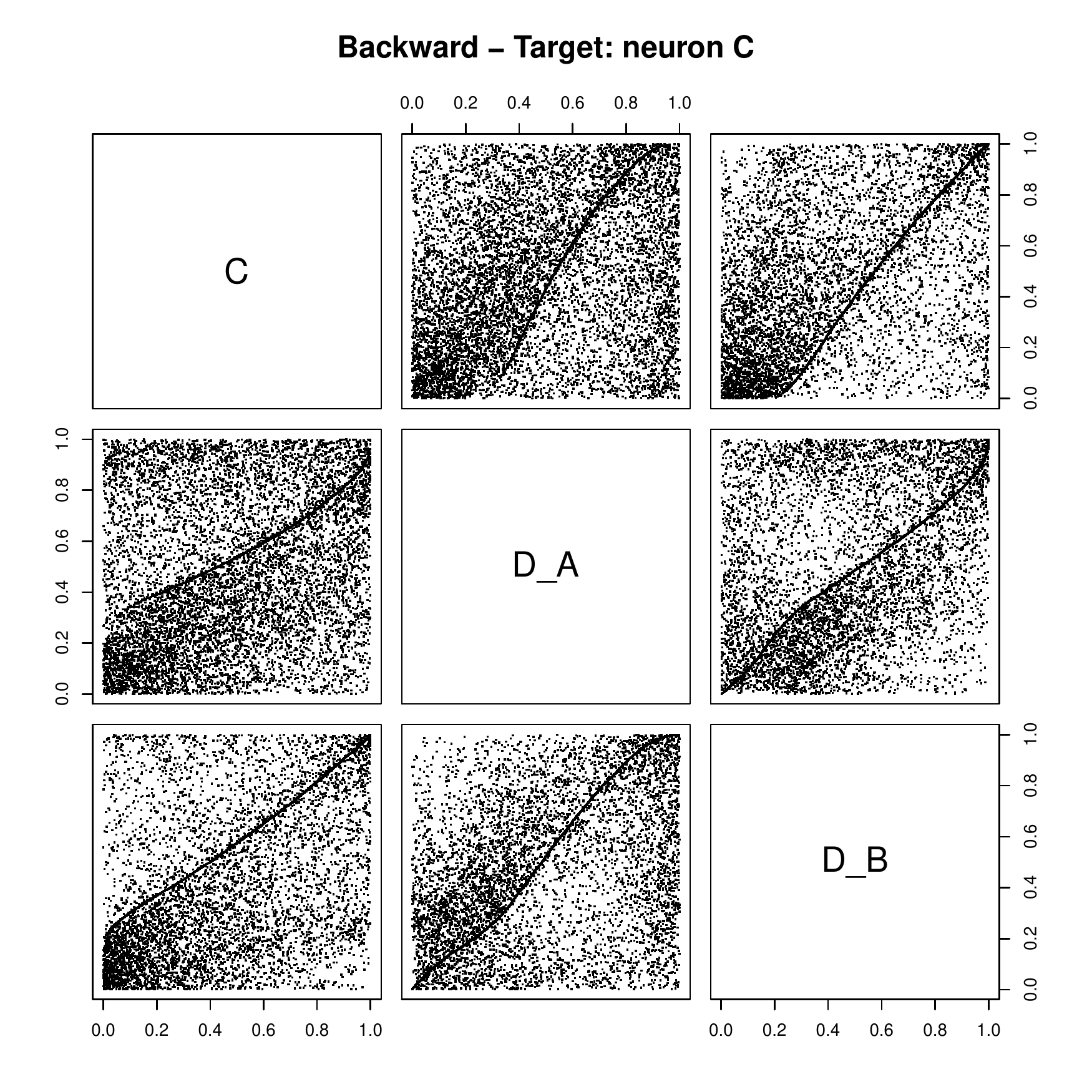}}
	\subfloat[][]
	{\includegraphics[width = 0.3\textwidth]{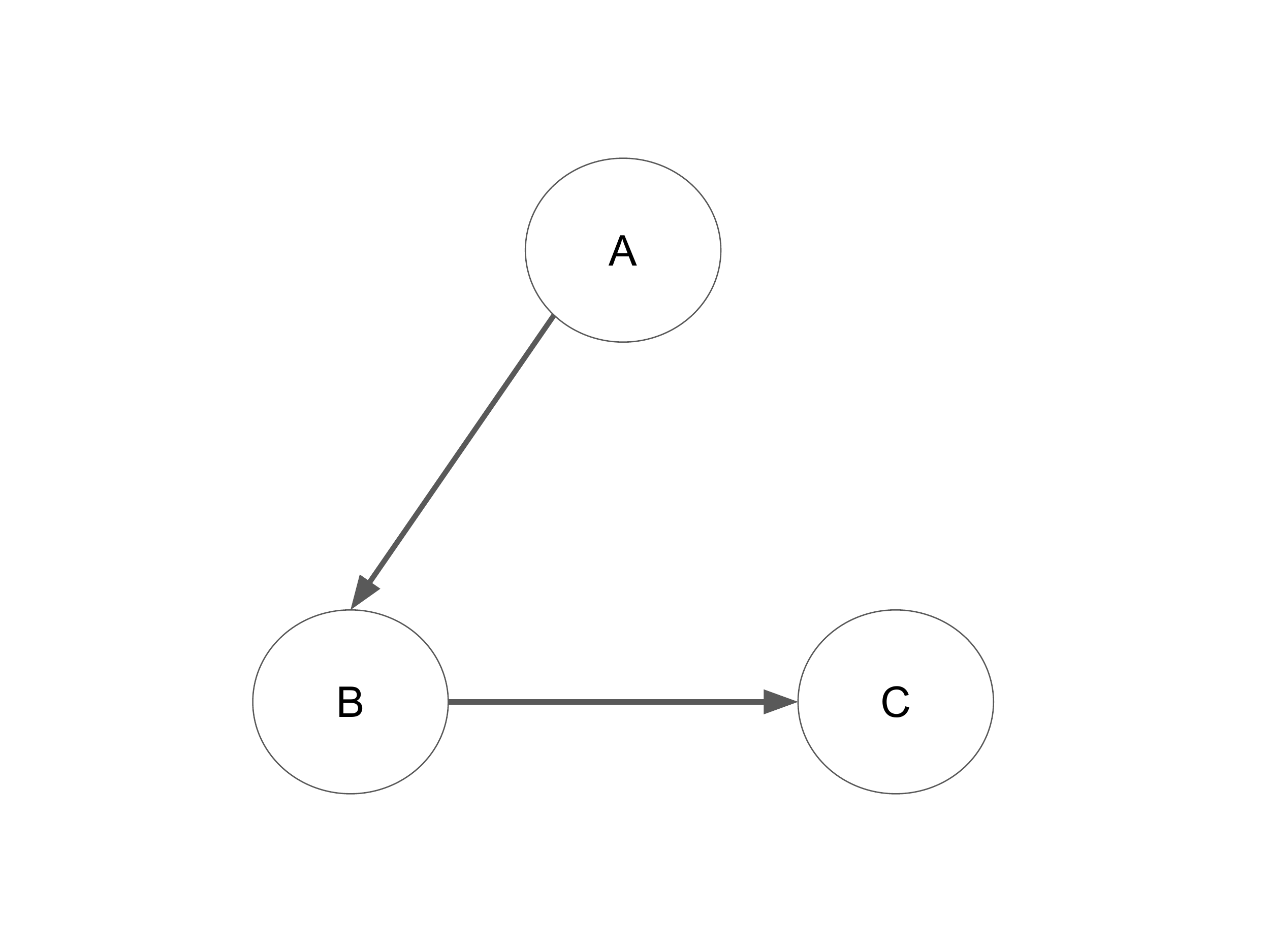}}
	\caption {Copula scatterplots for the spike trains and neural network used to generate them. By using the backward and the forward copulas a chain-like structure can be detected. Here the arrows represent excitatory connections with intensity $h=3$mV. }
	\label{fig:E29}
\end{figure}

Fig.~\ref{fig:E29} aims at illustrating the importance of the choice of the target neuron and of the use of forward/backward approach.  In panel (a) we use the forward approach and $A$ is the target neuron. The scatterplots are similar to those in Fig.~\ref{fig:E15}, suggesting the existence of direct links. Now, the crosscorrelograms plots dependencies between \acp{ISI} and forward times and it becomes difficult to guess their directions. The only information we can gather is that $A$ tends to fire first.

Choosing $C$ as the target in the backward approach  (see Fig.~\ref{fig:E29_C_BWD}), we observe lines similar to \textit{almost} straight lines. Panels coupling \acp{ISI} of $C$ with backward times of $B$ are the most similar to a straight line, indicating a direct link between $B$and $C$. Also Panels involving backward times of (B) and (A), i.e. two intervals of the same type, is quasi-linear. This indicates a direct link between the two neurons. On the contrary, the crosscorrelogram between $C$and the backward time of $A$ is more irregular. This result can be explained by the chain-like structure in Fig.~\ref{fig:E29} in which part of the signal arriving to $B$ is not sufficient to excite also $C$.  Note that the frequency of $C$ is higher than the one of $A$ (because of the induced jumps) and due to the presence of $B$ this can only be spotted backward, choosing the excited neuron as a target. 

\begin{figure}
	\centering
		\subfloat[][\label{fig:E31_A_FWD}] 
		{\includegraphics[width = 0.4\textwidth]{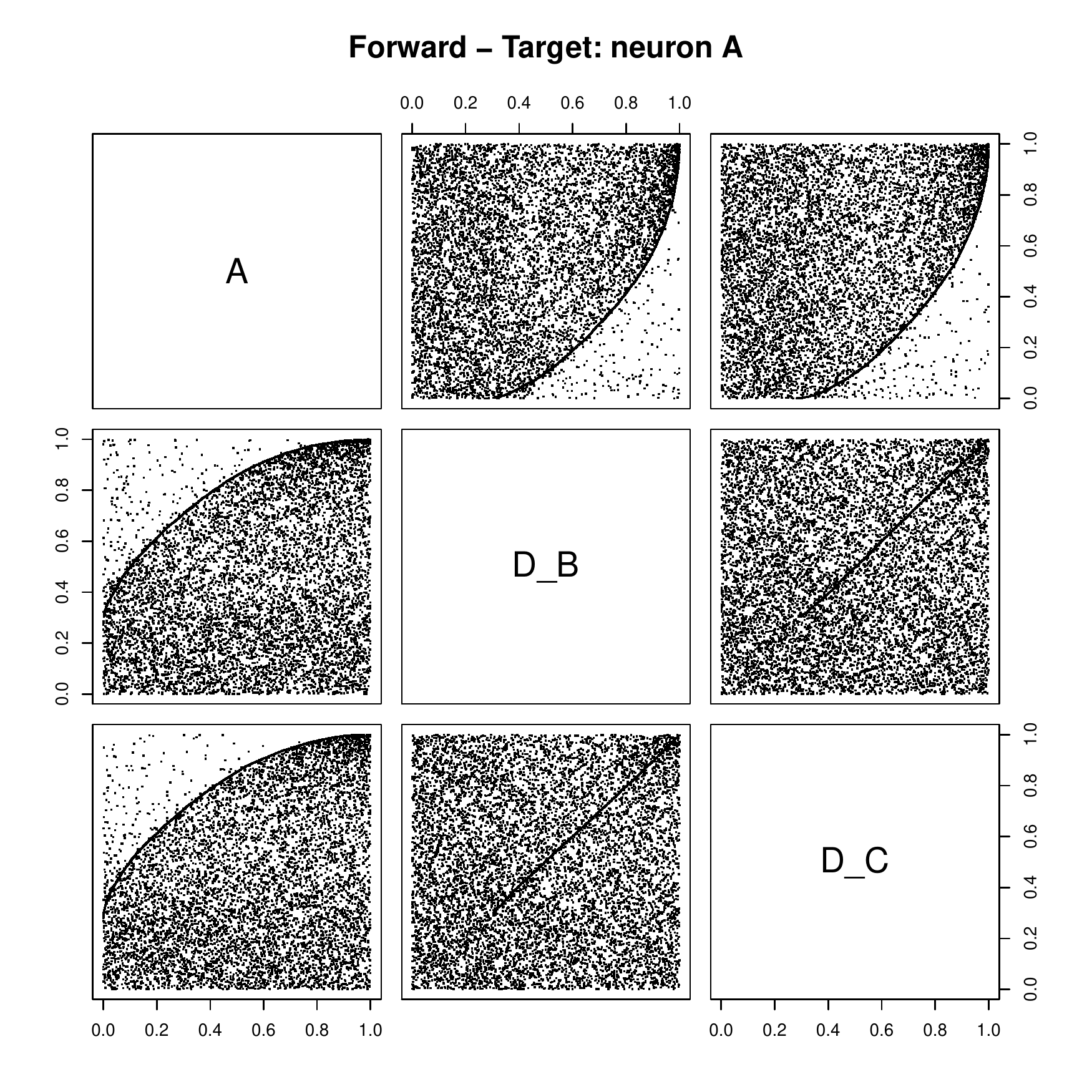}}
		\quad
		\subfloat[][\label{fig:E31_A_BWD}] 
		{\includegraphics[width = 0.4\textwidth]{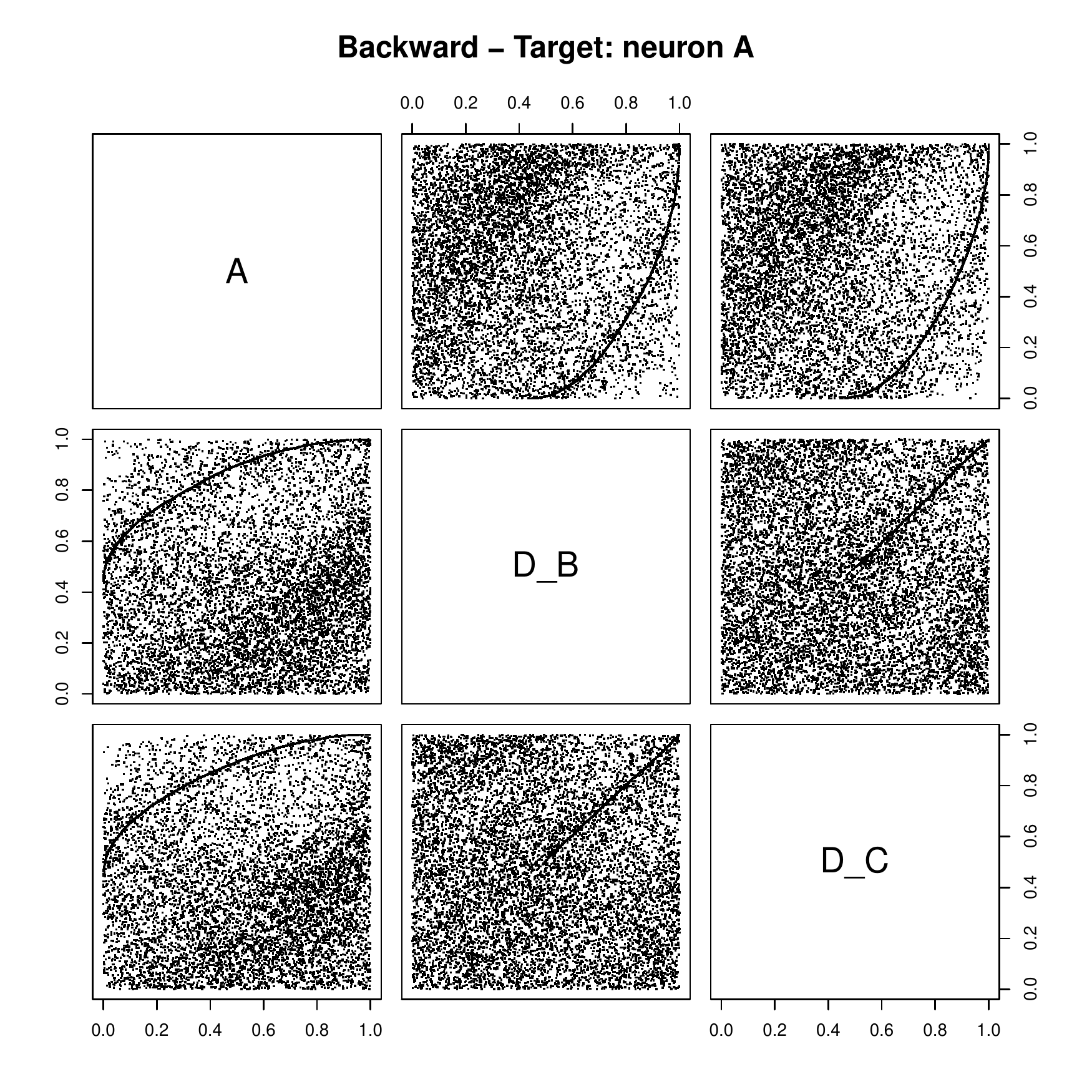}}
		\\
		\subfloat[][\label{fig:E31_B_BWD}] 
		{\includegraphics[width = 0.4\textwidth]{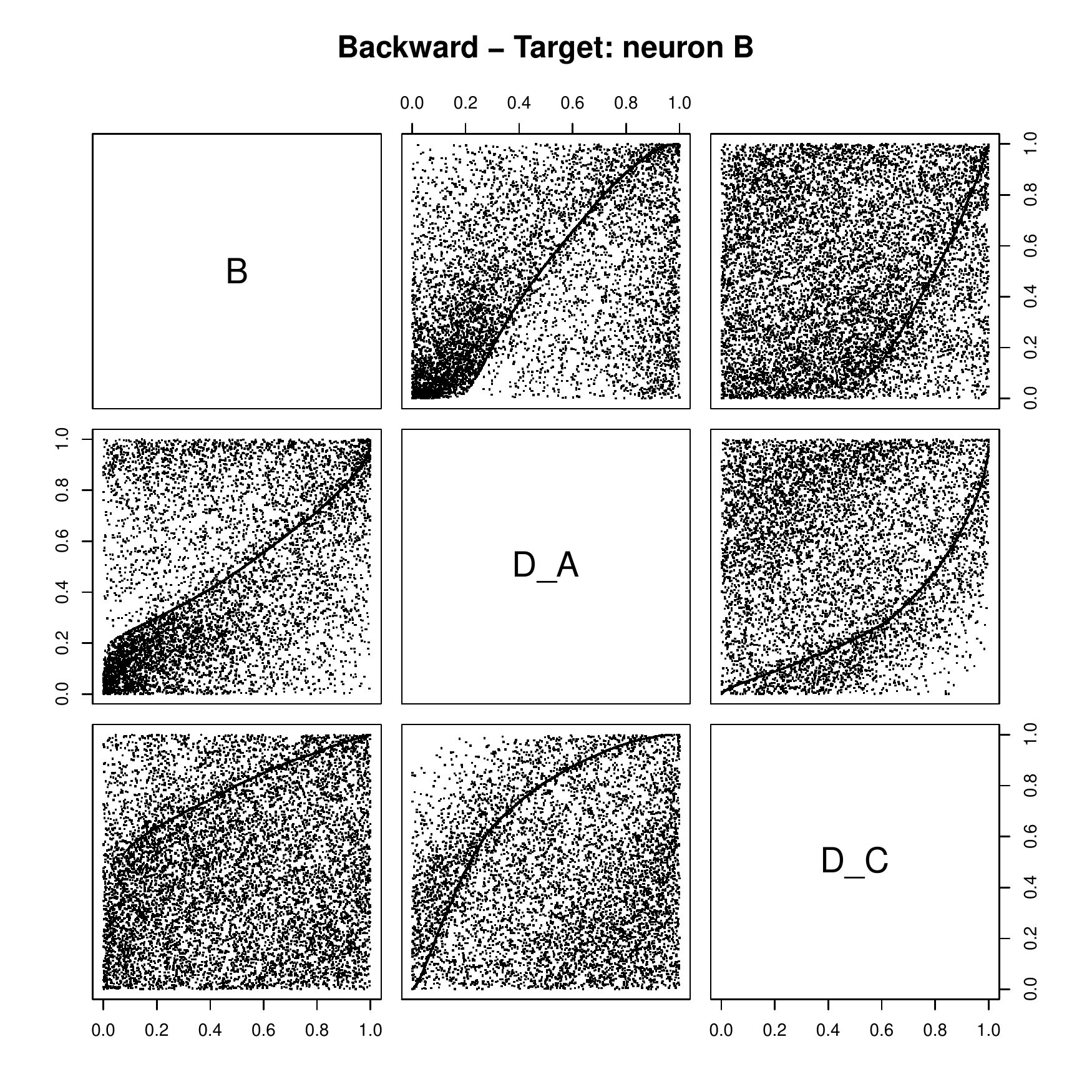}}
		\quad
		\subfloat[][\label{fig:E31_C_BWD}] 
		{\includegraphics[width = 0.4\textwidth]{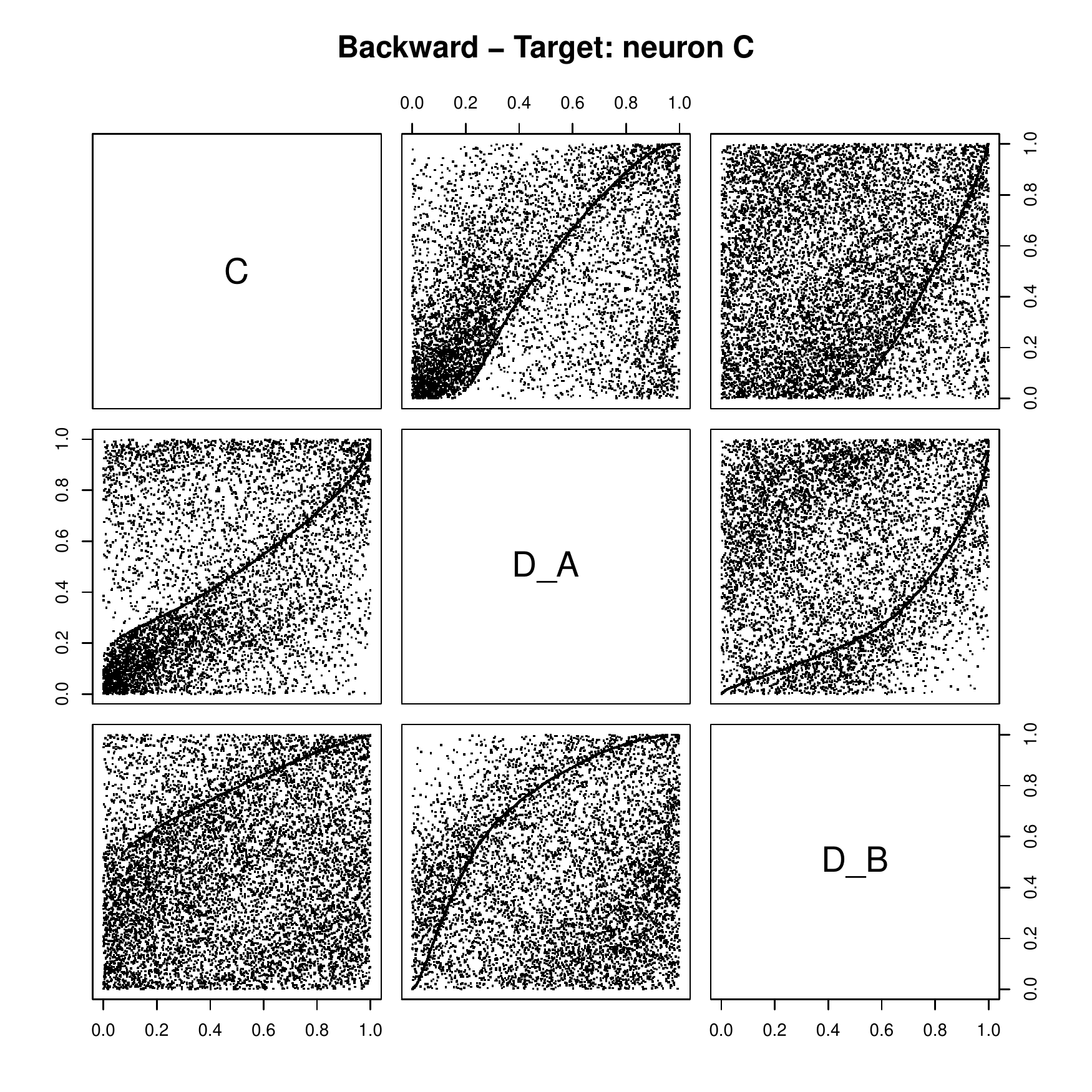}}
		\\
	\subfloat[][\label{fig:E31_g}]
	{\includegraphics[width = 0.3\textwidth]{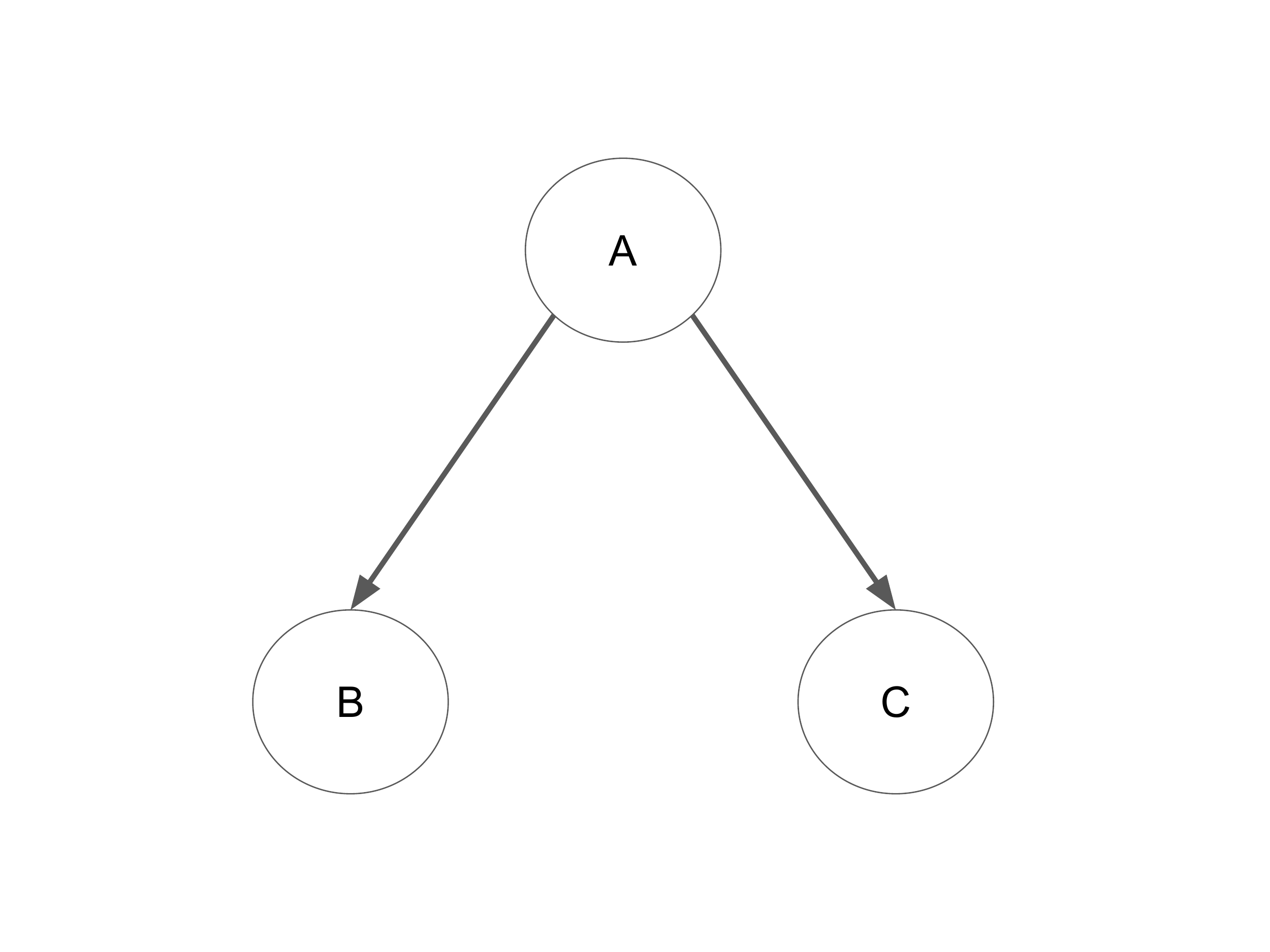}}
	\caption {Copula scatterplots for the spike trains and neural network used to generate them. The mono-directional excitatory connections can be spotted comparing copula scatterplots obtained by choosing different neurons as a target. Here the arrows represent excitatory connections with intensity $h=3$mV. }
	\label{fig:E31}
\end{figure}

A further illustration of the difference between backward and forward approaches is presented in the top panels of Fig.~\ref{fig:E31}.  Panel \subref{fig:E31_A_FWD} shows clearly the dependence of $B$ and $C$ from $A$, while panel \subref{fig:E31_A_BWD} is very noisy. Furthermore, Panel \subref{fig:E31_A_FWD} shows the existence of synchronous spikes of neurons (B) and (C). The presence on many asynchronous spikes for these neurons suggests the existence of a signal helping the spike of both of them at the same time.   
Moreover all the scatterplots in \subref{fig:E31_B_BWD} and \subref{fig:E31_C_BWD}, using different target neurons in the backward approach, look really similar. This observation suggest a similar input to neurons (B) and (C).
All this remarks, together with other figures not shown, are in agreement with the structure of the neural network in panel \subref{fig:E31_g}. This coincides with the model used to generate the samples.

\begin{figure}
	\centering
	\subfloat[][\label{fig:E34_C_BWD}] 
	{\includegraphics[width = 0.45\textwidth]{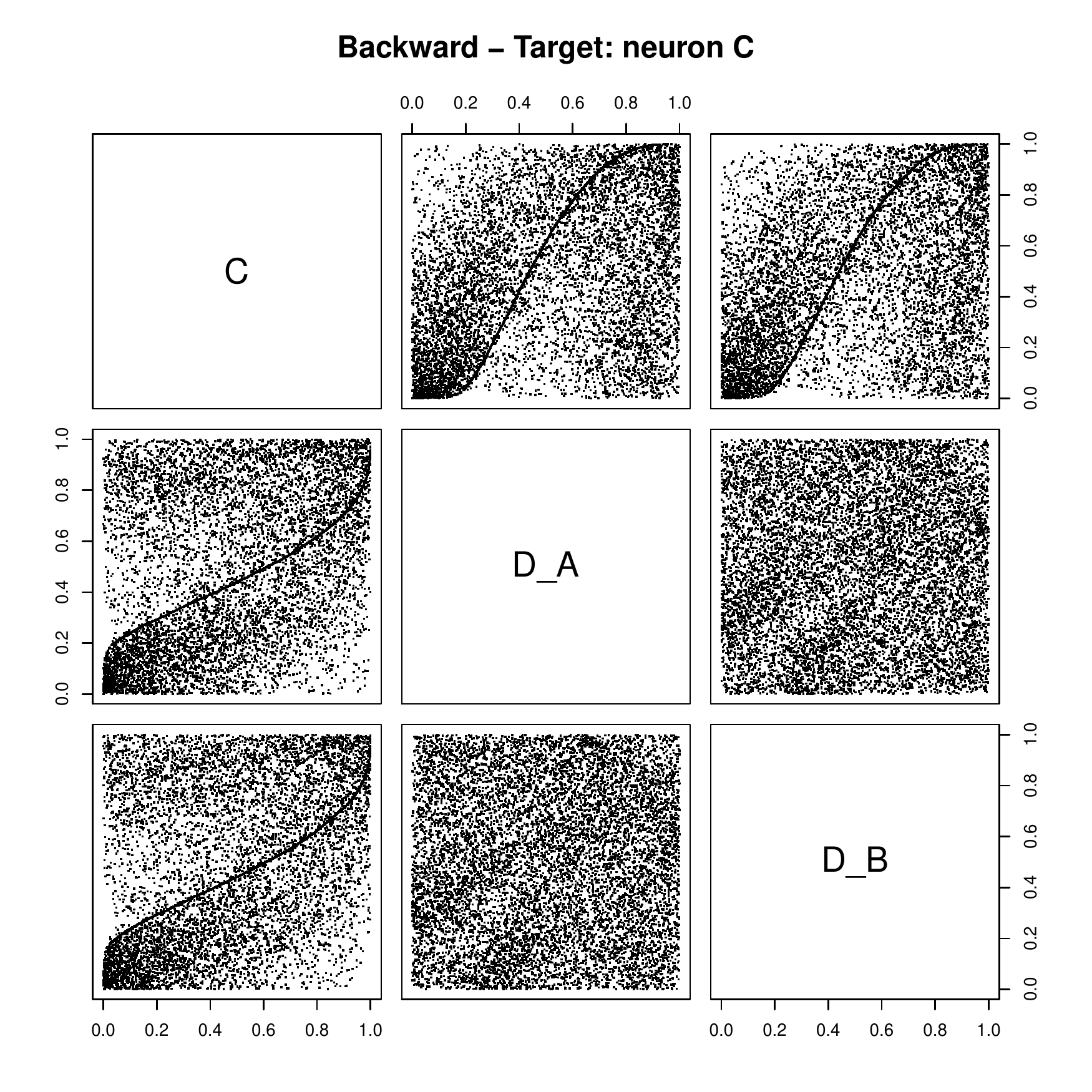}}
	\subfloat[][\label{fig:E34_g}]
	{\includegraphics[width = 0.45\textwidth]{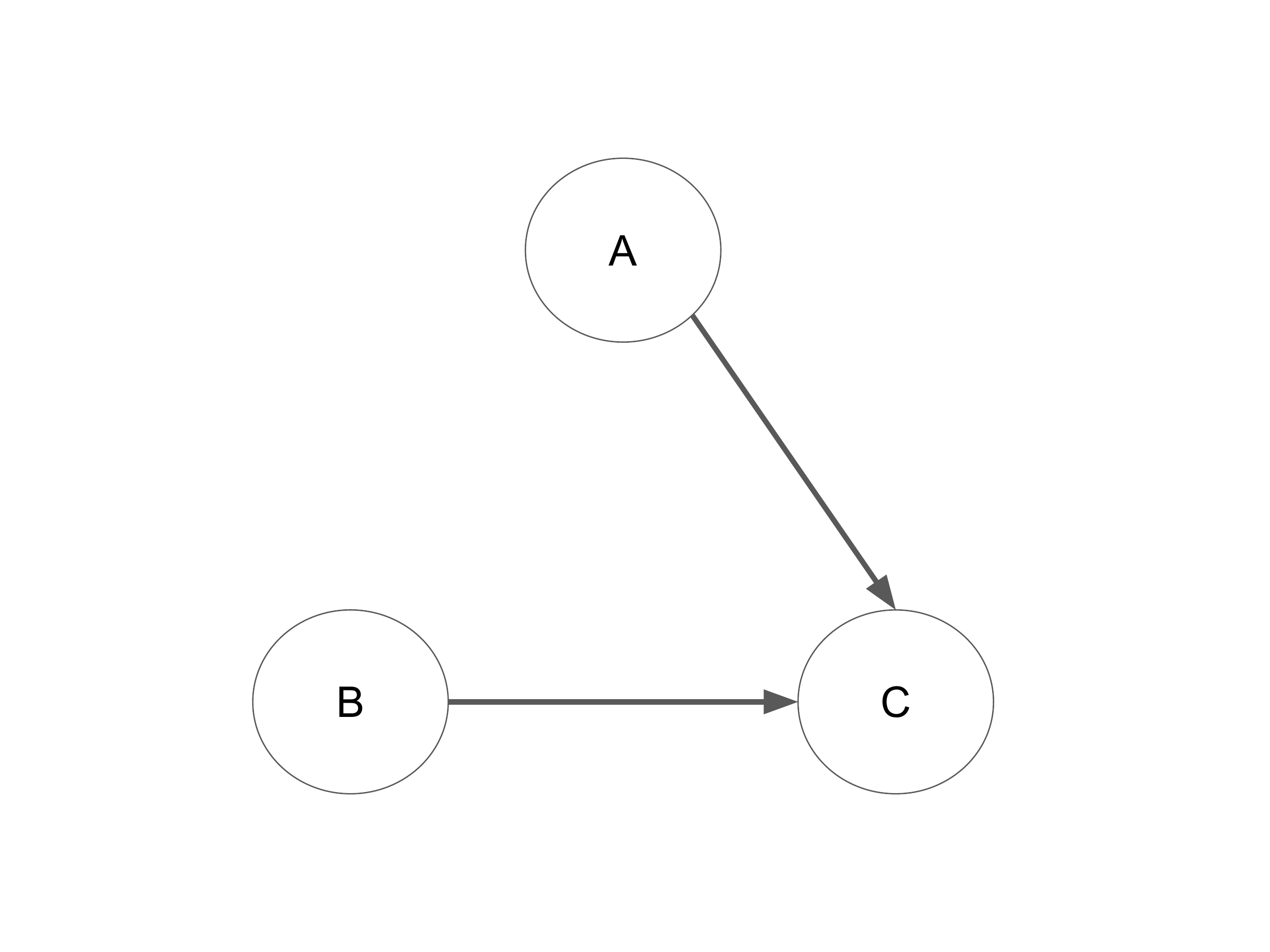}}
	\caption {Copula scatterplots for the spike trains and neural network used to generate them. Note how the independence of the neuron $A$ and $B$ is evident from the copula scatterplot. Here the arrows represent excitatory connections with intensity $h=3$mV. }
	\label{fig:E34} 
\end{figure}

Finally, Fig.~\ref{fig:E34}  shows how, choosing $C$ as a target, in the backward approach, we easily recognize the independence of neuron $A$ and $B$. The copula between their backward times is the independent copula. Moreover, the dependencies of $C$ from $A$ and $B$ corresponds to very similar scatterplots.  These observations are coherent with the structure used to generate the spike trains (Fig.~\ref{fig:E34_g}) that can be detected from the scatterplots.

We also considered examples in which negative jumps mimic the presence of inhibition. The corresponding scatterplots show an absence of points corresponding to the impossibility to observe spikes at certain times, depending from the other neurons activity. We do not report these figures for space reasons but the analysis can be prformed along lines similar to those illustrated for the excitatory case.

\section{Conclusions}\label{sec:Conclusions}
Pursuing on the research line proposed in \cite{sacerdote2012detecting} we use copulas to recognize dependencies between involved quantities. We consider here networks of three neurons but wishing to use scatterplots of the considered copulas we limited our study to bivariate copulas. The novelty of this paper is related with the joint use of Forward and Backward times together with \acp{ISI} to guess the structure of a neural network from observed spike trains. Through a set of examples we illustrated the usefulness of the proposed method. 
We showed that the presence of noisy scatterplots for the \acp{ISI} or for other pairs of intertimes is determined by the existence of indirect links. Furthermore, we showed that the direction of the links can be determined studying scatterplots. This aim requests a complete study involving the change the target neuron and the use of both forward and backward times between the neurons.\\

The paper aims at illustrating the usefulness of the copula approach. Copulas catch all the information present in the sample and we show here how to choose the random variables for their study. The present study considers only networks of three neurons. Extensions to the case of an higher number of neurons is theoretically possible but requests the use of many scatterplots to include different combinations between forward and backward times, as well as between different target neurons. Furthermore,  increasing the number of involved neurons we expect more noisy figures.
The analysis of experimental data will surely determine a major variety of features and new difficulties of interpretation.  However, the analysis of synthetic data helps to learn how to read sets of scatterplots, developing an ability that will become important to switch to the analysis of experimental data.




\section*{Acknowledgments}
This work was partially supported by INdAM-GNCS.

\bibliographystyle{model1-num-names}

\bibliography{MyBib}

\begin{acronym}
	\acro{eCDF}{empirical Cumulative Distribution Function}
	\acro{CDF}{Cumulative Distribution Function}
	\acro{ePSP}{excitatory Post-Synaptic Potential}
	\acro{FPT}{First-Passage Time}
	\acro{iid}{identical and independently distributed}
	\acro{iPSP}{inhibitory Post-Synaptic Potential}
	\acro{IF}{Integrate and Fire}
	\acro{ISI}{Inter-Spike Interval}
	\acro{LIF}{Leaky Integrate and Fire}
	\acro{MP}{Membrane Potential}
	\acro{NERVE}{Neural Environment for Random Variables Estimation}
	\acro{OU}{Ornstein-Uhlenbeck}
	\acro{PCC}{Pair-Copula Construction}
	\acro{PDF}{Probability Density Function}
	\acro{PSP}{Post-Synaptic Potential}
	\acro{RRW}{Randomized Random Walk}
	\acro{RV}{Random Variable}
	\acro{SDE}{Stochastic differential equation}
\end{acronym}







\end{document}